\documentclass[10pt,twocolumn,letterpaper]{article}

\usepackage{iccv}
\usepackage{times}
\usepackage{epsfig}
\usepackage{graphicx}
\usepackage{amsmath}
\usepackage{amssymb}
\usepackage{booktabs}
\usepackage{bm}
\usepackage{booktabs}
\usepackage{array}
\usepackage{algorithmic}
\usepackage{relsize}
\usepackage{subcaption}
\usepackage{lipsum}
\usepackage{multirow}
\usepackage{authblk}
\usepackage{cite} 
\usepackage{multicol,lipsum,environ}
\usepackage{float}

\usepackage[breaklinks=true,bookmarks=false]{hyperref}

\newcommand{\bx}{{\bf x}}
\newcommand{\bz}{{\bf z}}
\newcommand{\bI}{{\bf I}}
\newcommand{\bh}{{\bf h}}
\makeatletter
\newcommand\xleftrightarrow[2][]{%
	\ext@arrow 9999{\longleftrightarrowfill@}{#1}{#2}}
\newcommand\longleftrightarrowfill@{%
	\arrowfill@\leftarrow\relbar\rightarrow}
\makeatother

\newcolumntype{C}[1]{>{\centering\let\newline\\\arraybackslash\hspace{0pt}}m{#1}}

\iccvfinalcopy 
\def\iccvPaperID{****} %

\begin{document}

\title{DUAL-GLOW: Conditional Flow-Based Generative Model for Modality Transfer}

\author[1,2,3]{Haoliang Sun}
\author[1]{Ronak Mehta}
\author[1]{Hao H. Zhou}
\author[1]{Zhichun Huang}
\author[1]{\\Sterling C. Johnson}
\author[1]{Vivek Prabhakaran}
\author[1]{Vikas Singh}

\affil[ ]{\texttt{haolsun.cn@gmail.com, ronakrm@cs.wisc.edu, \{hzhou97,zhuang294,prabhakaran\}@wisc.edu, scj@medicine.wisc.edu, vsingh@biostat.wisc.edu}}
{

\affil[1]{University of Wisconsin-Madison, Madison, USA}
\affil[2]{Shandong University, Jinan, CN}
\affil[3]{Inception Institute of Artificial Intelligence, Abu Dhabi, UAE}
}

\maketitle
\ificcvfinal\thispagestyle{empty}\fi

\begin{abstract}
Positron emission tomography (PET) imaging is an
imaging modality for diagnosing a number of
neurological diseases. 
In contrast to Magnetic Resonance Imaging (MRI), PET is costly and involves injecting a radioactive substance into the patient.
Motivated by developments in modality transfer in vision, 
we study
the generation of certain types of PET images from MRI data. 
We derive new flow-based
generative models which we show perform
well in this small sample size regime 
(much smaller than dataset sizes available in standard vision tasks). 
Our formulation, DUAL-GLOW, is based on two invertible networks
and a relation network that maps the latent spaces to each other.
We discuss how given the prior distribution,
learning the conditional distribution of PET given the MRI image
reduces to obtaining the conditional distribution between the two latent codes w.r.t. the two image types.
We also extend our framework to leverage ``side'' information (or attributes) when available.
By controlling the PET generation through ``conditioning'' on age, our model is also
able to capture brain FDG-PET (hypometabolism) changes, as a function of age. 
We present experiments on the Alzheimer’s Disease Neuroimaging Initiative (ADNI) dataset with 826 subjects, and  obtain good performance in PET image synthesis, qualitatively and quantitatively better than recent works.

\end{abstract}

\vspace{-4mm}
\section{Introduction}
\begin{figure*}[t]
	\begin{center}
		\centerline{\includegraphics[width=0.8\linewidth]{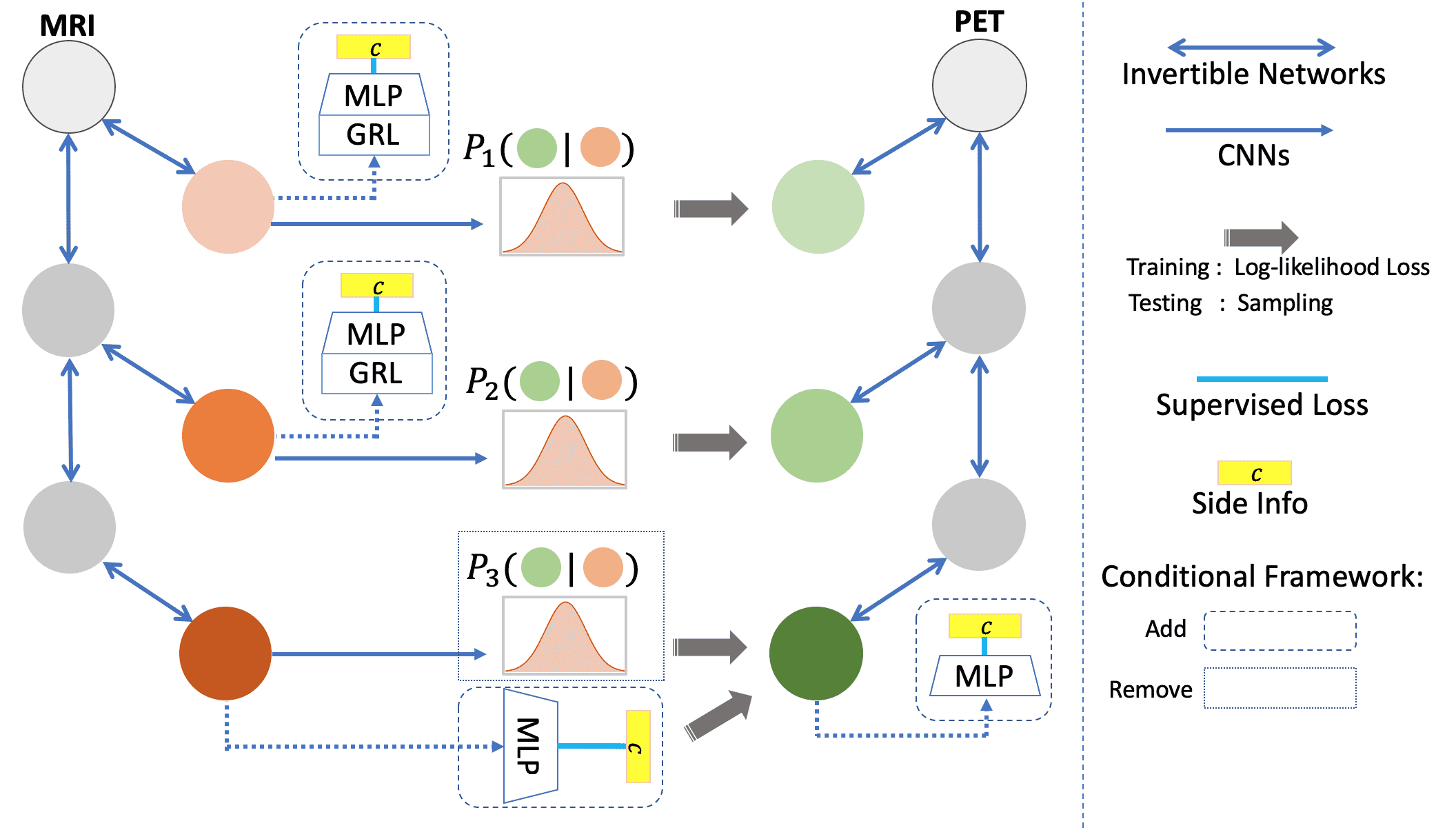}}
	\end{center}
	\vspace{-32pt}
	\caption{The DUAL-GLOW framework.
          For the conditional module,
          the dashed and dotted pieces are added and removed respectively.
          The colored circle represents the latent code whereas the gray one is the image or the intermediate output.}
	\label{fig:framework}
	\vspace{-10pt}
\end{figure*}

Positron Emission Tomography (PET) images provide a three-dimensional image volume
reflecting metabolic activity in the tissues, e.g., brain regions, 
which is a key imaging modality for a number of diseases (e.g., Dementia, Epilepsy, Head and Neck Cancer). 
Compared with Magnetic Resonance (MR) imaging, the typical PET imaging procedure usually involves radiotracer injection and
a high cost associated with specialized hardware and tools, logistics, and expertise.
Due to these factors, Magnetic Resonance (MR)
imaging is much more ubiquitous than PET imaging in both clinical and research settings. Clinically, PET imaging is often only considered much further down the pipeline, after information from other non-invasive approaches has been collected. It is not uncommon for many research studies to include MR images for {\em all}
subjects, and acquire specialized PET images only for a {\em smaller subset} of participants.

{\bf Other use cases.} Leaving aside the issue of disparity in costs between MR and PET,
it is not uncommon to find that due to a variety of reasons other than cost,
a (small or large) subset of individuals in a study have {\em one or more
image scans unavailable}. 
Finding ways to ``generate'' one type of imaging modality given another
is attracting a fair bit of interest in the community and a number of ideas have been presented \cite{pan2018synthesizing}.
Such a strategy, if effective, can increase the sample sizes available
for statistical analysis and possibly, even for training downstream
learning
models for diagnosis. 

{\bf Related Work.} Modality transfer can
be thought of ``style transfer'' \cite{ isola2017image,zhu2017unpaired,zhu2017toward,kim2017learning,wang2018perceptual, murez2018image, lin2018conditional, choi2018stargan, liu2017unsupervised, huang2018multimodal, lee2018diverse, gonzalez2018image, mo2018instanceaware, ma2018exemplar,wang2018mix} in the context of medical
images and
a number of interesting
results in this area have appeared \cite{li2014deep, kang2015prediction, wolterink2017deep,nie2017medical,maspero2018dose,han2017mr,pan2018synthesizing}.
Existing methods, mostly based on deep learning for modality transfer,
can be roughly divided into two categories: Auto-encoders and
Generative Adversarial Networks (GANs) \cite{goodfellow2014generative,karras2017progressive,brock2018large}.
Recall that auto-encoders are composed of two modules, encoder and decoder.
The encoder maps the input to a hidden code $h$,
and the decoder maps the hidden code to the output.
The model is trained by minimizing the loss in the output Euclidean space with standard norms ($\ell_1$, $\ell_2$).
A U-Net structure, introduced in \cite{ronneberger2015u},
is typically used for leveraging local and hierarchical information to achieve an
accurate reconstruction. Although the structure in auto-encoders is elegant with
reasonable efficiency and a number of authors have reported good performance \cite{nie2017medical,sikka2018mri},
constructions based on minimizing the $\ell_2$ loss often produce blurry outputs, as has been observed in 
\cite{pan2018synthesizing}. Partly due to these reasons, more recent works
have investigated other generative models. 
Recently, one of the prominent generative models in use today, GANs \cite{goodfellow2014generative},
has seen much success
in natural image synthesis \cite{brock2018large}, estimating the generative model via
an adversarial process.
Despite their success in generating \textit{sharp} realistic images, GANs usually suffer from
``mode collapse", that tends to produce limited sample variety \cite{che2016mode,arjovsky2017wasserstein}.
This issue is only compounded in medical images,
where the maximal mode may simply be attributed to anatomical structure shared by most subjects. Further,
sample sizes are often much smaller in medical imaging compared to computer vision,
which necessitates additional adjustments to the architecture and parameters, as we found in our experiments as well. 

{\bf Flow-based generative models.} Another family of methods,
flow-based generative models \cite{dinh2014nice, dinh2016density, kingma2018glow}, has been proposed for variational inference and natural image generation and have only recently begun to gain attention in the computer vision community. A (\textit{normalizing}) \textit{flow}, proposed in \cite{rezende2015variational}, uses a sequence of invertible mappings to build the transformation of a probability density to approximate a posterior distribution. The flow starts with an initial variable and maps it to a variable with a simple distribution (e.g., isotropic Gaussian) by repeatedly applying the change of variable rule, similar to the inference procedure
in an encoder network. For the image generation task, the initial variable is the real image with some unknown probability function. Designating a well-designed inference network, the flow will learn an accurate mapping after training. Because the flow-based model is invertible, the generation of synthetic images is straightforward
by sampling from the simple distribution and ``flowing" through the map in reverse.
Compared with other generative models and Autoregressive Models \cite{oord2016pixel},
flow-based methods allow tractable and accurate log-likelihood evaluation during the training process, while also providing an efficient and exact sampling from the simple prior distribution at test time.

{\bf Where is the gap?} While flow-based generative models have been successful
in image synthesis, it is challenging to leverage them directly for modality transfer.
It is difficult to apply existing flow-based methods
to our task due to the invertibility constraint in the inference network.
Apart from various technical issues, consider an intuitive example. 
Given an MRI, we should expect that there would be
many solutions of corresponding PET images, and \textit{vice versa}.
Ideally, we prefer the model to provide a conditional distribution
of the PET given an MRI -- such a conditional distribution can also be meaningfully
used when additional information about the subject is available. 

{\bf This work.} Motivated by the above considerations, we propose
a novel flow-based generative model, DUAL-GLOW,
for MRI-to-PET image generation. The value of our model includes explicit latent variable representations,
exact and efficient latent-variable inference, 
and the potential for memory and computation savings through constant network size.
Utilizing recent developments in flow-based generative models by \cite{kingma2018glow}, DUAL-GLOW is
composed of two invertible inference networks and a relation CNN network, as pictured in Figure \ref{fig:framework}.
We adopt the multi-scale architecture with spliting technique in \cite{dinh2016density}, which can significantly reduce the
computational cost and memory. The two inference networks are built to project MRI and PET into two
semantically meaningful latent spaces, respectively. 
The relation network is constructed to estimate the conditional distribution between paired latent codes. 
The foregoing properties of the DUAL-GLOW framework enable specific improvements in \textit{modality transfer}
from MRI to PET images. 
Sampling efficiency allows us to process and generate full 3D brain volumes.

{\bf Conditioning based on additional information.}
While the direct generation of PET from MRI has much practical utility, it is often also
the case that a single MRI could correspond to a very different PET image -- and which images are far more likely
can be resolved based on additional information, such as age or disease status. 
However, a challenge arises due to the high correlation
between the input MR image and side information:
traditional conditional frameworks \cite{mirza2014conditional, kingma2018glow} cannot effectively generate meaningful images in this setting. To
accurately account for this correlation, we propose a
new conditional framework, see Figure \ref{fig:framework}, where
two small discriminators (Multiple Layer Perceptron, MLP) are concatenated at the end of the
top inference networks to faithfully extract the side information contained in the images. The remaining
two discriminators concatenated at the left invertible inference network are combined with Gradient Reverse Layers (GRL),
proposed in \cite{ganin2014unsupervised}, to exclude the side information which
exists
in the latent codes except at the top-most layer.
After training, sampling from the conditional distribution allows the generation of diverse and meaningful PET images.
Extensive experiments show the efficiency of this exclusion architecture in the conditional framework for side information manipulation.

\textbf{Contributions.} This paper provides: 
\textbf{(1)} A novel flow-based generative model for modality transfer, DUAL-GLOW. 
\textbf{(2)} A complete end-to-end PET image generation from MRI for full three-dimensional volumes.
\textbf{(3)} A simple extension that enables {\em side condition manipulation} -- a practically
useful property that allows assessing change as a function of age, disease status, or other covariates.
\textbf{(4)} Extensive experimental analysis of the quality of PET images generated by DUAL-GLOW,
indicating the potential for direct application in practice to help in the clinical evaluation of Alzheimer's
disease (AD).

\section{Flow-based Generative Models}
We first briefly review flow-based generative models to help motivate and present our algorithm.
Flow based generative models, e.g., GLOW \cite{kingma2018glow}, typically deal with single image generation.
At a high level, these approaches set up the task as calculating the log-likelihood of an input image with an unknown distribution.
Because maximizing this log-likelihood is intractable, a \textit{flow} is set up to project the data into a new space where it is easy to compute, as summarized below. 

Let $\bx$ be an image represented as a high-dimensional random vector in the image space with an unknown true distribution $\bx \sim p^\ast(\bx)$. We collect an i.i.d. dataset $\mathcal{D}$ with samples $\{\bx^i\}_{i=1}^n$ and choose a model class $p_\theta(\bx)$ with parameters $\theta$. Our goal is to find parameters $\hat{\theta}$ that produces $p_{\hat{\theta}}(\bx)$ to best approximate $p^\ast(\bx)$. This is achieved through maximization of the log-likelihood:
	
\begin{equation}
\mathcal{L}(\mathcal{D}) = \frac{1}{n}\sum_{i=1}^n \log p_\theta(\bx^i). \label{eq:1}
\end{equation}
In typical flow-based generative models \cite{dinh2014nice,dinh2016density,kingma2018glow}, the \textit{generative process} for $\bx$ is defined in the following way:
\begin{align}
\bz \sim p_\theta(\bz), \quad 
\bx = g_\theta(\bz), \label{eq:2}
\end{align}
where $\bz$ is the latent variable and $p_\theta(\bz)$ has a (typically simple) tractable density, such as a spherical multivariate Gaussian distribution: $p_\theta(\bz) = \mathcal{N}(\bz; 0, \bI)$. The function $g_\theta(\cdot)$ may correspond to a rich function class, but is invertible such that given a sample $\bx$, latent-variable inference is done by $\bz = f_\theta(\bx) = {g_\theta}^{-1}(\bx)$. For brevity, we will omit subscript $\theta$ from $f_\theta$ and $g_\theta$.

We focus on functions where $f$ is composed of a sequence of invertible transformations: $f = f_k \circ \cdots\circ f_2 \circ f_1$, where the relationship between $\bx$ and $\bz$ can be written as:
\begin{equation}
\bx \xleftrightarrow{f_1} \bh_1 \xleftrightarrow{f_2} \bh_2 \cdot\cdot\cdot  \xleftrightarrow{f_k} \bz.
\label{eq:flows}
\end{equation}

Such a sequence of invertible transformations is also called a (normalizing) {\it flow} \cite{rezende2015variational}. Under the change of variables rule through \eqref{eq:2}, the log probability density function of the model \eqref{eq:1} given a sample $\bx$ can be written as:
\begin{align}
\log p_\theta(\bx) &= \log p_\theta(\bz) + \log |det(d\bz/d\bx)|\\
&= \log p_\theta(\bz) + \sum_{i=1}^k \log |det(d\bh_i/d\bh_{i-1})|
\end{align}
where we define $\bh_0 = \bx$ and $\bh_k = \bz$ for conciseness. The scalar value $\log |det(d\bh_i/d\bh_{i-1})|$ is the logarithm of the absolute value of the determinant of the Jacobian matrix $(d\bh_i/d\bh_{i-1})$, also called the log-determinant. While it may look difficult,
this value can be simple to compute for certain choices of transformations, as previous explored in \cite{dinh2014nice}. For the transformations $\{f_i\}_{i=1}^k$ which characterizes the flow, there are several typical settings that result in invertible functions, including actnorms, invertible $1\times 1$ convolutions, and affine coupling layers \cite{kingma2018glow}. Here we use affine coupling layers, discussed in further detail shortly.
For more, details regarding these mappings we refer the reader to existing literature on flow-based models, including GLOW \cite{kingma2018glow}.

%
%
%

\begin{figure*}[]
	\centering
	\begin{minipage}{.7\textwidth}
		\vspace{4mm}
		\centering
		\includegraphics[width=1.03\linewidth]{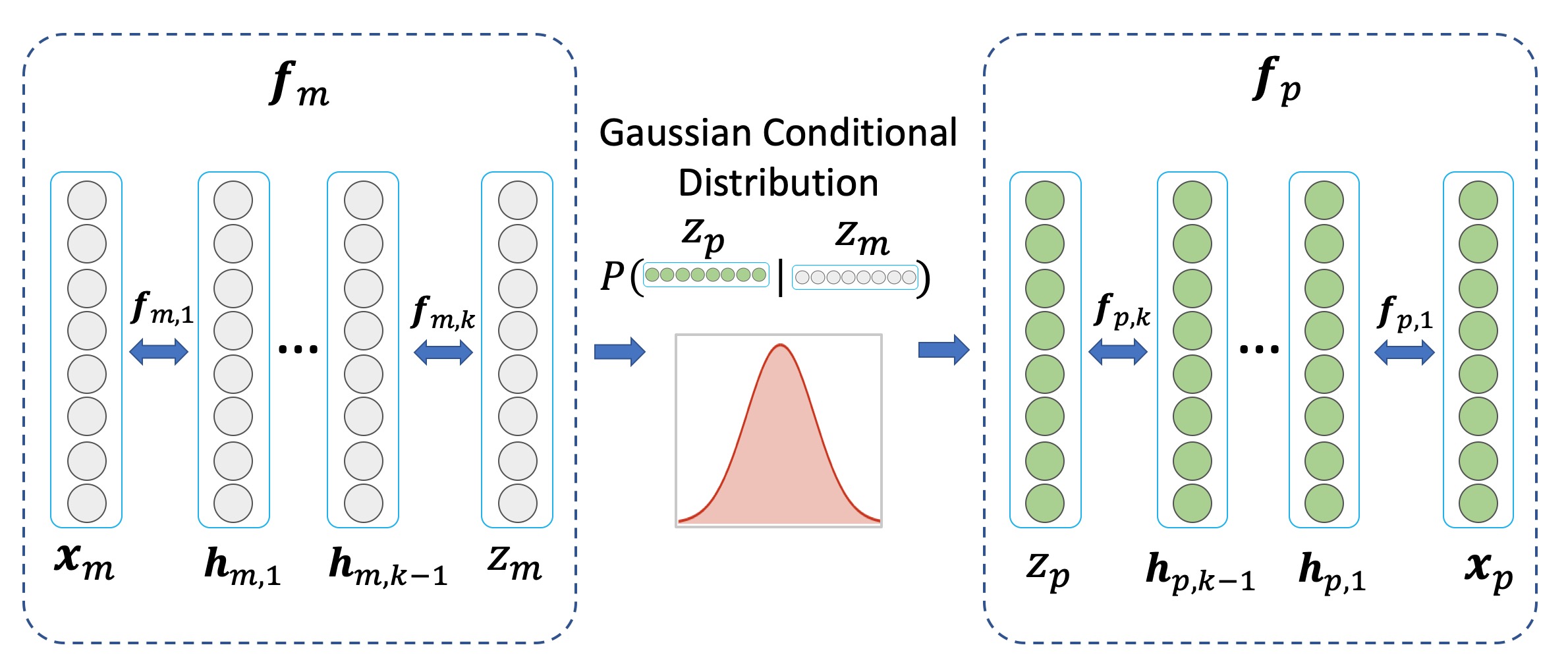}
		\vspace{-10pt}
		\captionof{figure}{DUAL-GLOW for image generation.}
		\label{fig:dual-g}
	\vspace{-5pt}
	\end{minipage}%
	\begin{minipage}{.3\textwidth}
		\centering
		\includegraphics[width=.8\linewidth]{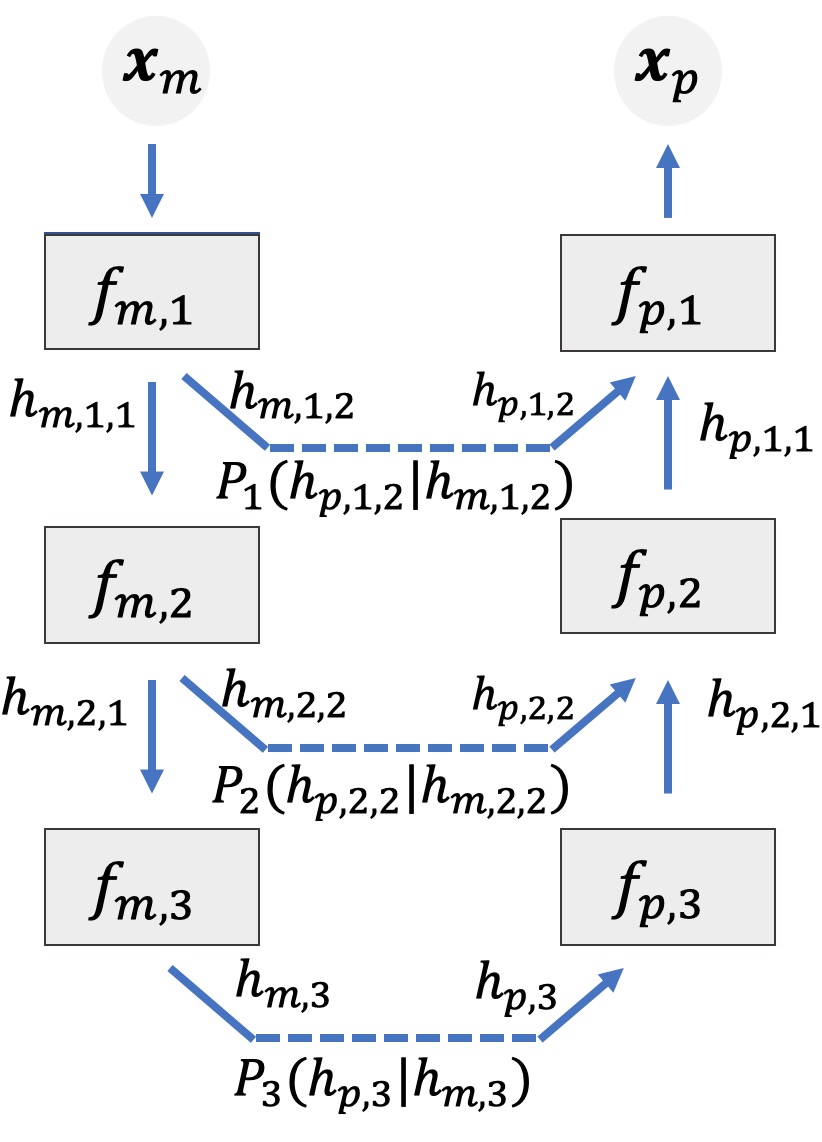}
		\vspace{-1pt}
		\captionof{figure}{Spliting.}
		\label{fig:split}
	\vspace{-10pt}
	\end{minipage}
\end{figure*}

\section{Deriving DUAL-GLOW}
In this section, we present our DUAL-GLOW framework for inter-modality transfer. We first
discuss the derivation of the conditional distribution of a PET image given an MR image and then
provide strategies for efficient calculation of its log-likelihood.
Then, we introduce the construction of the invertible flow and show the calculation for
the Jacobian matrix. Next, we build the hierarchical architecture for our DUAL-GLOW framework,
which greatly reduces the
computational cost compared to a flat structure.
Finally, the conditional structure for side information manipulation is derived with additional discriminators. 
 
\noindent{\bf Log-Likelihood of the conditional Distribution.} Let the data
corresponding to the MR and PET images be denoted as $\mathcal{D}_m$ and $\mathcal{D}_p$.
From a dataset $\mathcal{D}_m = \{\bx_m^i\}_{i=1}^{n}$, we are interested in generating images which have the same properties as images in the dataset
  $\mathcal{D}_p = \{\bx_p^i\}_{i=1}^{n}$.
In our DUAL-GLOW model,
we assume that there exists a flow-based invertible function $f_p$ which maps
the PET image $\bx_p$ to $\bz_p = f_p(\bx_p)$ and a flow-based invertible function $f_m$ which maps
the MR image $\bx_m$ to $\bz_m = f_m(\bx_m)$. The latent variables $\bz_p$ and $\bz_m$ help set up a conditional probability $p_\theta(\bz_p|\bz_m)$, given by
\begin{equation}
\label{eq:cond}
p_\theta(\bz_p|\bz_m) = \mathcal{N}(\bz_p; \mu_\theta(\bz_m), \sigma_\theta(\bz_m))
\end{equation}

The full mapping composed of $f_p$, $f_m$, $\mu_\theta$ and $\sigma_\theta$ formulates our DUAL-GLOW framework:
\begin{equation}
\bx_m \xleftrightarrow{f_m} \bz_m \xrightarrow{\mu_\theta, \sigma_\theta} \bz_p \xleftrightarrow{{f_p}^{-1}} \bx_p,
\end{equation}
see Figure \ref{fig:dual-g}. The invertible functions $f_p$ and $f_m$ are designed as flow-based invertible functions. The mean function $\mu_\theta$ and the
covariance function $\sigma_\theta$ for $p_\theta(\bz_p|\bz_m)$ are assumed to be specified by neural networks. 
In this generating process, our goal is to maximize the log conditional probability $p_\theta(\bx_p|\bx_m)$. By the change of variable rule, we have that
\begin{align}
\log& p_\theta(\bx_p|\bx_m) = \log \left(p_\theta(\bx_p, \bx_m) / p_\theta(\bx_m)\right) \\
& = \log p_\theta(\bz_p| \bz_m) + \log \left(\frac{|det(d (\bz_p, \bz_m) /d (\bx_p, \bx_m))| }{|det(d\bz_m / d\bx_m)|} \right) \label{eq:log_joint}\\
& =  \log p_\theta(\bz_p| \bz_m)  + \log\left(|det(d\bz_p / d\bx_p)|\right). \label{eq:log_cond}
\end{align}
Note that the Jacobian $d (\bz_p, \bz_m) /d (\bx_p, \bx_m)$ in \eqref{eq:log_joint} is, in fact, a block matrix 
\begin{equation}
\label{eq:block}
\frac{d (\bz_p, \bz_m)}{d (\bx_p, \bx_m)}  = \left[ \begin{array}{cc} d\bz_p / d\bx_p & 0 \\ 0 & d\bz_m / d\bx_m \end{array}\right].
\end{equation}
Recall that calculating the deteriminant of such a matrix is straightforward (see \cite{silvester2000determinants}), which leads directly to \eqref{eq:log_cond}.

Without any regularization, maximizing such a conditional probability can 
make the optimization hard.
Therefore, we may
add a regularizer by controlling the marginal distribution $p_\theta(\bz_m)$, which leads to our objective function
\vspace{-7pt}
\begin{align}
\max_{f_m,f_p,\mu_\theta,\sigma_\theta}  &\log p_\theta(\bz_p| \bz_m)  +\log(|det(\frac{d\bz_p}{d\bx_p}  )|)  + \lambda  \log p_\theta(\bx_m) 
\nonumber \\
&=\log p_\theta({f_p}(\bx_p)| f_m(\bx_m))  + \log(|det(\frac{d\bz_p}{d\bx_p})|) \label{eq:3} \nonumber  \\
& \quad + \lambda \log (p_\theta(f_m(\bx_m)) |det(\frac{d\bz_m}{d\bx_m} )|),
\end{align}
where $\lambda$ is a hyperparameter, $p_\theta(\bz_m) = \mathcal{N}(\bz_m; 0, \bI)$ and $p_\theta(\bz_p| \bz_m) =  \mathcal{N}(\bz_p; \mu_\theta(\bz_m), \sigma_\theta(\bz_m))$.

Interestingly, compared to GLOW, our model does {\bf not} introduce much additional complexity in computation. Let us see why. 
First, the marginal distribution $p_\theta(\bz)$ in GLOW is replaced by $p_\theta(\bz_p|\bz_m)$ and $p_\theta(\bz_m)$, which still has a
simple and tractable density. Second, instead of one flow-based invertible function in GLOW, our DUAL-GLOW
has two flow-based invertible functions $f_p$, $f_m$. Those functions are setup in parallel based on \eqref{eq:3}, extending the model size by a constant factor.


%

\noindent{\bf Flow-based Invertible Functions.}
In our work, we use an affine coupling layer to design the flows for the invertible functions $f_p$ and $f_m$. Before proceeding to the details, we omit subscripts $p$ and $m$ to simplify notations in this subsection. The invertible function $f$ is composed of a sequence of transformations $f = f_k \circ \cdots \circ f_2 \circ f_1$, as introduced in \eqref{eq:flows}.
In DUAL-GLOW, $\{f_i\}_{i=1}^k$ are designed by using the affine coupling layer \cite{dinh2016density} following these equations:
\begin{equation}
\begin{aligned}
&\bh_i  =  f_i(\bh_{i-1}) \Leftrightarrow \\ & 
\begin{cases}
\bh_{i; 1:d_1} = \bh_{i-1; 1:d_1}\\
\begin{aligned}
\bh_{i; d_1+1:d_1+d_2}=&(\bh_{i-1;  d_1+1:d_1+d_2} \odot \, \exp(\textrm{s}(\bh_{i-1; 1:d_1}))\\& + \; \; \textrm{t}(\bh_{i-1; 1:d_1}),
\end{aligned}
\end{cases}
\end{aligned}
\end{equation}
where $\, \odot \,$ denotes element-wise multiplication, $\bh_i \in R^{d_1+d_2}$, $\bh_{i; 1:d_1}$ the first $d_1$ dimensions of $\bh_i$, and $\bh_{i; d_1+1:d_1+d_2}$ the
remaining $d_2$ dimensions of $\bh_i$. The functions $s(\cdot)$ and $t(\cdot)$ are nonlinear transformations where it makes
sense to use deep convolutional neural networks (DCNNs). This construction makes the function $f$ invertible. To see this, we can easily write the inverse function ${f_i}^{-1}$ for $f_i$ as
\begin{equation}
\begin{aligned}
&\bh_{i-1} = (f_i)^{-1}(\bh_i) \Leftrightarrow \\ &  
\begin{cases}
\bh_{i-1; 1:d_1} = \bh_{i; 1:d_1}\\
\begin{aligned}
\bh_{i-1; d_1+1:d_1+d_2} =&(\bh_{i; d_1+1:d_1+d_2} - \textrm{t}(\bh_{i; 1:d_1}))
\\& \; \odot  \; \exp(-\textrm{s}(\bh_{i; 1:d_1})).
\end{aligned}
\end{cases}
\end{aligned}
\end{equation}

%

In addition to invertibility, this structure also tells us that the $\log(|det(d\bz_p / d\bx_p)|)$ term in our objective \eqref{eq:3} has a simple and tractable form. Computing the Jacobian, we have:
\begin{equation}
\frac{\partial f_i(\bh_{i-1})}{\partial \bh_{i-1}} = \left[ \begin{array}{cc} 
\textrm{I}_{1:d_1} & 0 \\ \frac{\partial f_{i; d_1+1:d_1+d_2}}{\partial \bh_{i-1; 1:d_1}}  & \;\; \textrm{diag}\big( \exp \big(\textrm{s}(\bh_{i-1; 1:d_1}) \big) \big)
\end{array}   \right],
\end{equation} 
where $I_{1:d_1} \in R^{d_1\times d_1}$ is an identity matrix. Therefore, 
\begin{equation*}\begin{aligned}
\log(|det(d\bz_p &/ d\bx_p)|)  = \sum_{i=1}^k \log(|det(d\bh_i / d\bh_{i-1})|) \\
&=  \sum_{i=1}^k \log(|det(\textrm{diag}\big( \exp \big(\textrm{s}(\bh_{i-1; 1:d_1}) \big) \big))|) \\
& =  \sum_{i=1}^k \log \left|\exp \left(\sum_{j=1}^{d_1} \textrm{s}(\bh_{i-1; j}) \right) \right|\\
\end{aligned}\end{equation*}
which can be computed easily and efficiently, requiring no on-the-fly matrix inversions \cite{kingma2018glow}.

\noindent{\bf Efficiency from Hierarchical Structure.}
The flow $f = f_k \circ \cdots f_2 \circ f_1$ can be viewed as a hierarchical structure. For the two datasets $\mathcal{D}_m = \{\bx_m^i\}_{i=1}^{n}$ and $\mathcal{D}_p = \{\bx_p^i\}_{i=1}^{n}$, it is computationally expensive to make all features of all samples go through the entire flow. Following implementation strategies in
previous flow-based models, we use the splitting technique to speed up DUAL-GLOW in practice, see Figure \ref{fig:split}. When a sample $\bx$ reaches the $i$-th transformation $f_i$ in the flow as $\bh_{i-1}$, we split $\bh_{i-1}$ in two parts $\bh_{i-1, 1}$ and $\bh_{i-1, 2}$, and take only one part $\bh_{i-1, 1}$ through $f_i$ to become $\bh_i = f_i(\bh_{i-1, 1})$. The other part $\bh_{i-1, 2}$ is
taken out
from the flow without further transformation. Finally, all those split parts $\{\bh_{i, 2}\}_{i=1}^{k-1}$ and the top-most $\bh_{k}$ are concatenated
together to form $\bz$. By using this splitting technique in the flow hierarchy, the part leaving the flow ``early'' goes through fewer transformations. As discussed in GLOW
and previous flow-based models, each transformation $f_i$ is usually rich enough that splitting saves computation
without losing much quality in practice. We provide the computational complexity in the appendix. Additionally, this hierarchical representation enables a more succinct extension to allow side information manipulation.

\noindent{\bf How to condition based on side information?}
As stated above, additional covariates should influence the PET image we generate, even with a very similar MRI.
A key assumption in many conditional side information frameworks is that these two inputs (the input MR and the covariate)
are independent of each other. Clearly, however, there exists a high correlation between MRI and side information such as age or gender or disease status.
In order to effectively incorporate this into our model, it is necessary to disentangle the side information from the intrinsic properties encoded in the latent representation $\bz_m$ of the MR image. 

Let $c$ denote the side information, typically a high-level semantic label (age, sex, disease status, genotype). In this case, we expect that the effect of this side information would be at a high level in relation to individual image voxels. As such, we expect that only the highest level of DUAL-GLOW should be affected by this. The latent variables $\bz_p$ should be conditioned on side variable $c$ and $\bz_m = \{\bh_{i, 2}\}_{i=1}^{k}$ except $\bh_{k}$. Thus, we can rewrite the conditional probability in \eqref{eq:3} by adding $c$: 
\begin{equation}
\begin{aligned}\label{eq:cond_frame}
\max_{f_m,f_p,\mu_\theta,\sigma_\theta} & \log p_\theta(\bz_p| \bz^\prime_m, c) \\ & +\log(|det(\frac{d\bz_p}{d\bx_p})|)  + \lambda  \log p_\theta(\bx_m),
\end{aligned}
\end{equation}
where $\bz^\prime_m = \{\bh_{i, 2}\}_{i=1}^{k-1}$ is independent on $c$, and 
\begin{equation}p_\theta(\bz_p|\bz^\prime_m, c) = \mathcal{N}(\bz_p; \mu_\theta(\bz^\prime_m, c), \sigma_\theta(\bz^\prime_m, c)).
\end{equation}
To disentangle the latent representation $\bz^\prime_m$ and exclude the side information in $\bz^\prime_m$, we leverage the well-designed conditional framework composed of both flow and discriminators. Specifically, the condition framework tries to exclude the side information from $\{\bh_{i, 2}\}_{i=1}^{k-1}$ and keep it in $\bh_{k}$ at the top level during training time. To achieve this, we concatenate a simple discriminator for each $\{f_i\}_{i=1}^{k-1}$ and add a Gradient Reversal Layer (GRL), introduced in \cite{ganin2014unsupervised}, at the beginning of the network. These classifiers are used for distinguishing the side information in a supervised way. The GRL acts as the identity function during the forward-propagation and reverses the gradient in back-propagation. Therefore, minimizing the classification loss in these classifiers is equivalent to pushing the model to exclude the information gained by this side information, leading to the exclusive representation $\bz^\prime_m = \{\bh_{i, 2}\}_{i=1}^{k-1}$. We also add a classifier \textit{without} GRL at the top level of $f_m, f_p$ that explicitly preserves this side information at the highest level.

Finally, the objective is the log-likelihood loss in \eqref{eq:cond_frame} plus the
classification losses, which can be jointly optimized by the popular optimizer AdaMax \cite{kingma2014adam}.
The gradient is calculated in a memory efficient way inspired by \cite{chen2016training}.
After training the conditional framework, we achieve PET image generation influenced {\bf both} by MRI and side information.

\section{Experiments}

\begin{figure}[b]
	\centering
	\includegraphics[width=\columnwidth]{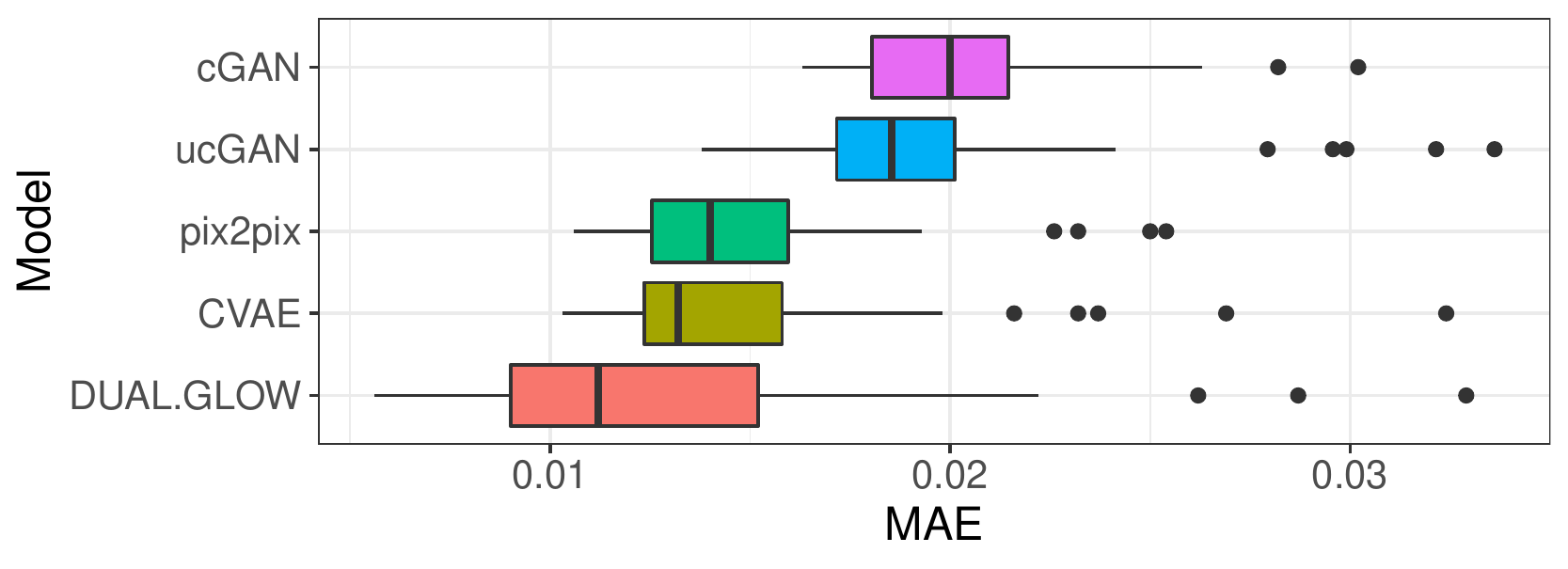}
	\caption{\label{fig:box_plot} Box plot of MAE metrics for different methods.}
\end{figure}

\begin{figure*}[t]
	\centering
	\begin{tabular}{c@{\hspace{3mm}}|c@{\hspace{3mm}}|c}
		
		\begin{tabular}{c@{\hspace{1mm}}c@{\hspace{1mm}}c@{\hspace{1mm}}}
			MRI & GroundTruth & Synthetic\\
			\begin{subfigure}{1.5cm}
				\centering\includegraphics[width=1.5cm]{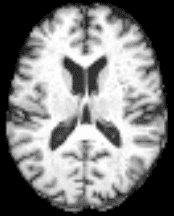}
			\end{subfigure}&
			\begin{subfigure}{1.5cm}
				\centering\includegraphics[width=1.5cm]{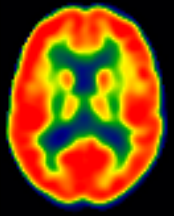}
			\end{subfigure}&
			\begin{subfigure}{1.5cm}
				\centering\includegraphics[width=1.5cm]{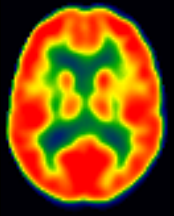}
			\end{subfigure} \\
			
			\begin{subfigure}{1.5cm}
				\centering\includegraphics[width=1.5cm]{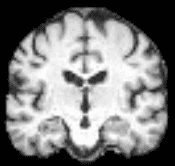}
			\end{subfigure}&
			\begin{subfigure}{1.5cm}
				\centering\includegraphics[width=1.5cm]{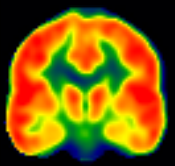}
			\end{subfigure}&
			\begin{subfigure}{1.5cm}
				\centering\includegraphics[width=1.5cm]{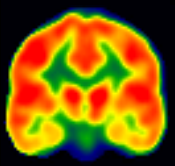}
			\end{subfigure} \\
			
			\begin{subfigure}{1.5cm}
				\centering\includegraphics[width=1.5cm]{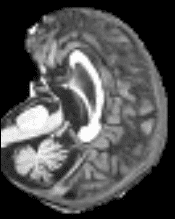}
			\end{subfigure}&
			\begin{subfigure}{1.5cm}
				\centering\includegraphics[width=1.5cm]{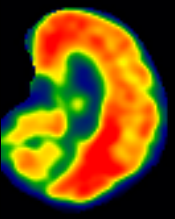}
			\end{subfigure}&
			\begin{subfigure}{1.5cm}
				\centering\includegraphics[width=1.5cm]{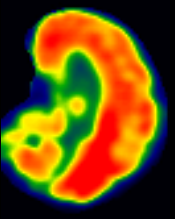}
			\end{subfigure}\\
			
		\end{tabular} &
		
		\begin{tabular}{c@{\hspace{1mm}}c@{\hspace{1mm}}c@{\hspace{1mm}}}
			MRI & GroundTruth & Synthetic\\
			
			\begin{subfigure}{1.5cm}
				\centering\includegraphics[width=1.5cm]{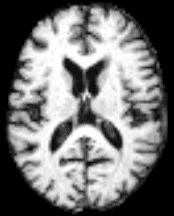}
			\end{subfigure}&
			\begin{subfigure}{1.5cm}
				\centering\includegraphics[width=1.5cm]{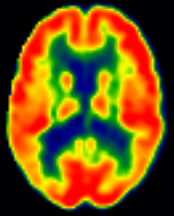}
			\end{subfigure}&
			\begin{subfigure}{1.5cm}
				\centering\includegraphics[width=1.5cm]{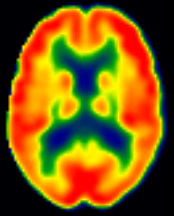}
			\end{subfigure} \\
			
			\begin{subfigure}{1.5cm}
				\centering\includegraphics[width=1.5cm]{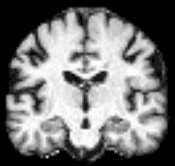}
			\end{subfigure}&
			\begin{subfigure}{1.5cm}
				\centering\includegraphics[width=1.5cm]{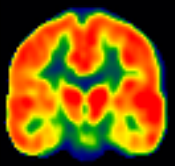}
			\end{subfigure}&
			\begin{subfigure}{1.5cm}
				\centering\includegraphics[width=1.5cm]{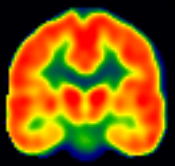}
			\end{subfigure} \\
			
			\begin{subfigure}{1.5cm}
				\centering\includegraphics[width=1.5cm]{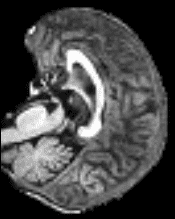}
			\end{subfigure}&
			\begin{subfigure}{1.5cm}
				\centering\includegraphics[width=1.5cm]{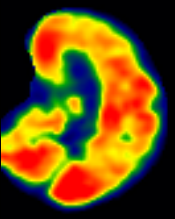}
			\end{subfigure}&
			\begin{subfigure}{1.5cm}
				\centering\includegraphics[width=1.5cm]{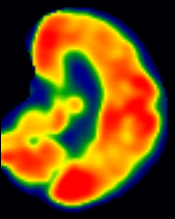}
			\end{subfigure}\\
			
		\end{tabular} &

		\begin{tabular}{c@{\hspace{1mm}}c@{\hspace{1mm}}c@{\hspace{1mm}}}
			MRI & GroundTruth & Synthetic\\
			
			\begin{subfigure}{1.5cm}
				\centering\includegraphics[width=1.5cm]{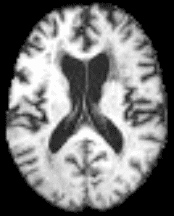}
			\end{subfigure}&
			\begin{subfigure}{1.5cm}
				\centering\includegraphics[width=1.5cm]{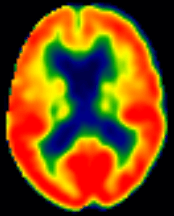}
			\end{subfigure}&
			\begin{subfigure}{1.5cm}
				\centering\includegraphics[width=1.5cm]{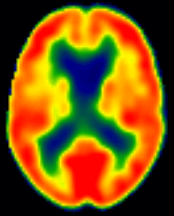}
			\end{subfigure} \\
			
			\begin{subfigure}{1.5cm}
				\centering\includegraphics[width=1.5cm]{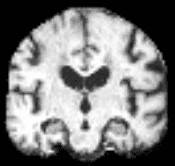}
			\end{subfigure}&
			\begin{subfigure}{1.5cm}
				\centering\includegraphics[width=1.5cm]{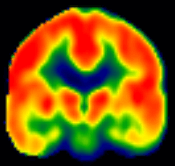}
			\end{subfigure}&
			\begin{subfigure}{1.5cm}
				\centering\includegraphics[width=1.5cm]{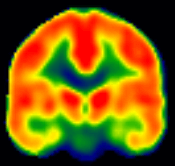}
			\end{subfigure} \\
			
			\begin{subfigure}{1.5cm}
				\centering\includegraphics[width=1.5cm]{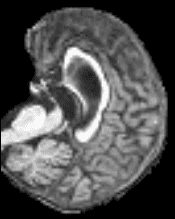}
			\end{subfigure}&
			\begin{subfigure}{1.5cm}
				\centering\includegraphics[width=1.5cm]{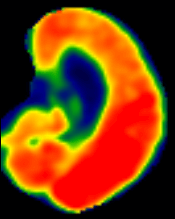}
			\end{subfigure}&
			\begin{subfigure}{1.5cm}
				\centering\includegraphics[width=1.5cm]{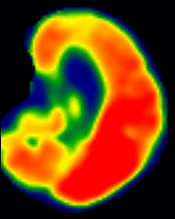}
			\end{subfigure}\\
			
		\end{tabular}\\

	\end{tabular}
	
	\vspace{-3pt}
	\caption{\label{fig:samples-cn} {\bf Synthetic images are meaningful for subjects in both
			extremes of disease spectrum}. Left: CN. Middle: MCI. Right: AD. The generated PET images show consistency of hypometabolism (less red, more yellow) with the ground truth image. (Best viewed in color; montages shown in the appendix).}
	\vspace{-5pt}
\end{figure*}

We evaluate the model's efficacy on the ADNI dataset both against ground truth images and for downstream applications.
We conduct extensive quantitative experiments which
show that DUAL-GLOW outperforms the baseline method consistently. Our generated PET images show desirable clinically meaningful properties
which is relevant for their potential use in Alzheimer's Disease diagnosis.
The conditional framework also shows promise in tracking hypometabolism as a function of age.

\subsection{ADNI Dataset}
\noindent\textbf{Data.} The Alzheimer's Disease Neuroimaging Initiative (ADNI)
provides a large database of studies
directly aimed at understanding the 
development and pathology of Alzheimer's Disease. Subjects are diagnosed as cognitively normal (CN), significant memory concern (SMC), early mild cognitive impairment (EMCI), mild cognitive impairment (MCI), late mild cognitive impairment (LMCI) or having Alzheimer's Disease (AD).
FDG-PET and T1-weighted MRIs were
obtained from ADNI,
and pairs were constructed
by matching images
with the same subject ID 
and similar acquisition dates.

\noindent\textbf{Preprocessing.} Images were processed using SPM12 \cite{ashburner2014spm12}. First, PET images were aligned to the paired MRI using coregistration. Next, MR images were nonlinearly mapped
to the MNI152 template.
Finally, PET images were mapped to the standard MNI space using the same forward warping identified in the MR segmentation step.
Voxel size was fixed for all volumes to 
$1.5\times 1.5 \times 1.5 \:  mm^3$,
and the final volume size obtained for 
both MR and PET images was $64 \times 96 \times 64$.
Through this workflow, we finally
obtain 806 MRI/PET clean pairs. The demographics of the dataset are provided in the appendix. 
In the
following experiments, we randomly select 726 subjects as the training data and the remaining 80 as testing within a 10-fold evaluation scheme.

\noindent\textbf{Framework Details.}
The DUAL-GLOW architecture outlined above was trained using Nvidia V100 GPUs with Tensorflow. There are $4$ ``levels" in our invertible network, each containing $16$ affine coupling layers. The nonlinear operators $\textrm{s}(\cdot)$ and $\textrm{t}(\cdot)$ are small networks with three 3D convolutional layers. For the hierarchical correction learning network, we split the hidden codes of the output of the first three modules in the invertible network and design four 3D convolutional networks for all latent codes. For the conditional framework case, we concatenate the five discriminators to
the tail of all four levels of the MRI inference network and the top-most level of the PET inference network. The GRL is added between the inference network and the first three discriminators. The hyperparameter $\lambda$ is the regularizer and
set to $0.001$. For all classification losses, we set the weight to $0.01$. 
The model was trained using the AdamMax optimizer with an initial learning rate set to $0.001$ and exponential decay rates $0.9$ for the moment estimates. We train the model for $90$ epochs. Our implementation is available at \url{https://github.com/haolsun/dual-glow}.

\begin{figure*}[t]
	\centering
	\begin{tabular}{c@{\hspace{5mm}} c@{\hspace{6mm}}c@{\hspace{2mm}}c@{\hspace{2mm}}c@{\hspace{2mm}}c@{\hspace{2mm}}c@{\hspace{2mm}}c}
		& MRI & 50 & 60 & 70 & 80 & 90 & 100 \\
		AD &
		\begin{subfigure}{1.9cm}
			\centering\includegraphics[width=1.9cm]{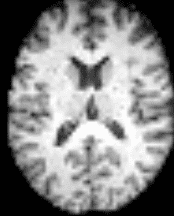}
		\end{subfigure} &
		\begin{subfigure}{1.9cm}
			\centering\includegraphics[width=1.9cm]{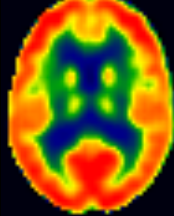}
		\end{subfigure}&
		\begin{subfigure}{1.9cm}
			\centering\includegraphics[width=1.9cm]{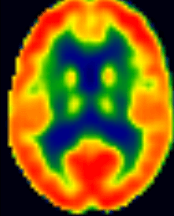}
		\end{subfigure}&
		\begin{subfigure}{1.9cm}
			\centering\includegraphics[width=1.9cm]{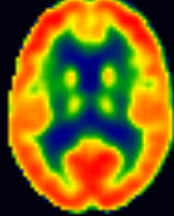}
		\end{subfigure}&
		
		\begin{subfigure}{1.9cm}
			\centering\includegraphics[width=1.9cm]{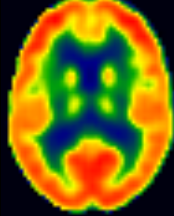}
		\end{subfigure}&
		\begin{subfigure}{1.9cm}
			\centering\includegraphics[width=1.9cm]{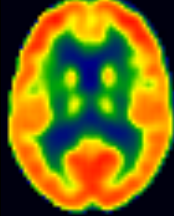}
		\end{subfigure}&
		\begin{subfigure}{1.9cm}
			\centering\includegraphics[width=1.9cm]{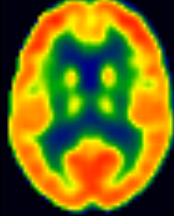}
		\end{subfigure}\\ \\
		

		CN &
		\begin{subfigure}{1.9cm}
			\centering\includegraphics[width=1.9cm]{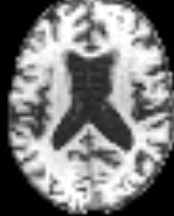}
		\end{subfigure} &
		\begin{subfigure}{1.9cm}
			\centering\includegraphics[width=1.9cm]{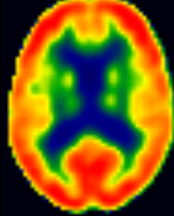}
		\end{subfigure}&
		\begin{subfigure}{1.9cm}
			\centering\includegraphics[width=1.9cm]{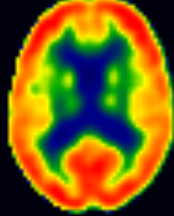}
		\end{subfigure}&
		\begin{subfigure}{1.9cm}
			\centering\includegraphics[width=1.9cm]{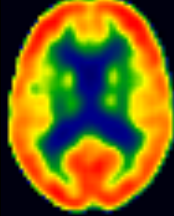}
		\end{subfigure}&
		
		\begin{subfigure}{1.9cm}
			\centering\includegraphics[width=1.9cm]{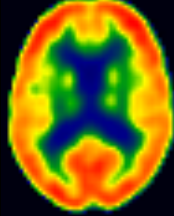}
		\end{subfigure}&
		\begin{subfigure}{1.9cm}
			\centering\includegraphics[width=1.9cm]{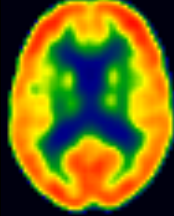}
		\end{subfigure}&
		\begin{subfigure}{1.9cm}
			\centering\includegraphics[width=1.9cm]{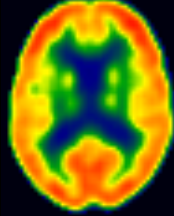}
		\end{subfigure}\\
		
	\end{tabular}
	\vspace{-4pt}
	\caption{\label{fig:cond_samples} {\bf Conditioning on age should yield generated images that
			show increased hypometabolism with age.} These are representative
		results from our PET generation as a function of age.
		As we scan left to right, we indeed see a decrease in metabolism (less red, more yellow)
		which is completely consistent with what we would expect in aging. (Best viewed in color; montages shown in the appendix).}

	\vspace{-12pt}
\end{figure*}

\subsection{\label{sec: quality}Generated versus Ground Truth consistency}

We begin our model evaluation by comparing outputs from our model to 4 state-of-the-art methods used previously for similar image generation tasks, conditional GANs (cGANs) \cite{nie2017medical}, cGANs with U-Net architecture (UcGAN) \cite{ronneberger2015u}, Conditional VAE (C-VAE) \cite{esser2018variational, sohn2015learning}, pix2pix \cite{isola2017image}.  Additional experimental setup details are in the appendix. We compare using commonly-used quantitative measures computed over the held out testing data. These include Mean Absolute Error (MAE), Correlation Coefficients (CorCoef), Peak Signal-to-Noise Ratio (PSNR), and Structure Similarity Index (SSIM). For Cor\_Coef, PSNR and SSIM, higher values indicate better generation of PET images.
For MAE, the lower the value, the better is the generation. 
As seen in Table \ref{etb1} and Figure \ref{fig:box_plot}, our model competes favorably against other methods.

\begin{table*}[]
	\begin{center}
		\caption{\label{etb1} Quantitative comparison results on 10-fold cross-validation.}
		\vspace{-3mm}
			\begin{tabular}{|c | c |c |  c | c | c|}
				\hline
				\footnotesize{METHOD} & cGAN & UcGAN & C-VAE & pix2pix  &  DUAL-GLOW \\
				\hline \hline 
				CorCoef & $0.956$       & $0.963$    &$\bm{0.980}$ &$0.967$&  $0.975$      \\ \hline 
				PSNR     & $27.37\pm2.07$ & $27.84\pm1.23$ &$28.69\pm2.06$&$27.54\pm1.95$& $\bm{29.56} \pm \bm{2.66}$\\	\hline 
				SSIM     & $0.761\pm0.08$ & $0.780\pm0.06$ & $0.817\pm0.06$ &$0.783\pm0.05$&  $\bm{0.898} \pm \bm{0.06}$ \\
				\hline
		\end{tabular}
	\end{center}
\vspace{-17pt}
\end{table*}

Figure \ref{fig:samples-cn} shows test images generated after 90 epochs for Cognitively Normal
and Alzheimer's Disease individuals.
Qualitatively, not only is the model able to accurately reconstruct large scale anatomical structures
but it is also able to identify
minute, sharp boundaries between gray matter and white matter.
While here we focus on data
from individuals with a clear progression of
Alzheimer's disease from those who are clearly 
cognitively healthy, in preclinical cohorts
where disease signal may be weak, accurately constructing
finer-grained details may be critical
in identifying those who may be undergoing neurodegeneration due to dementia. More results are shown in the appendix.

\subsection{Scientific Evaluation of Generation}
\vspace{-5pt}
As we saw above, our method is able to 
learn the modality mapping from MRI to PET. 
However, often image acquisition is used as a means to an end:
typically towards disease diagnosis or informed preventative care.
While the generated images may seem computationally and visually coherent,
it is important that the images generated add some value towards these
downstream analyses.

We also evaluate the {\bf generated} PET images for disease prediction and classification. Using the AAL atlas, we obtain all 116 ROIs via atlas-based segmentation \cite{wu2007optimum} and use the mean intensity of each as image features.
A support vector machine (SVM) is trained with the standard RBF kernel (e.g., see \cite{hinrichs2009mkl}) to predict binary disease status (Normal, EMCI, SMC vs. MCI, LMCI, AD) for both the ground truth and the
generated images. The SVM trained on generated images achieves comparable accuracy and false positive/negative rates (Table \ref{etb2}), suggesting that the generated images contain sufficient discriminative signal for disease diagnosis.

\begin{table}[]
	\begin{center}
		\caption{\label{etb2}  Validation on the ground truth and synthetic images for the AD/CN classification.}
		\vspace{-8pt}
		\begin{tabular}{lC{2cm}  C{2cm} }
			\hline
			& Ground Truth    & Synthetic \\
			\hline                                                  
			Accuracy &$94\%$ & $91\%$\\
			False Negative Rate & $6\%$ & $6\%$ \\
			False Positive Rate & $0\%$ & $3\%$ \\
			\hline 
		\end{tabular}
	\end{center}
	\vspace{-15pt}
\end{table}

\begin{figure}[t]
	\centering
	\includegraphics[width=0.95\columnwidth]{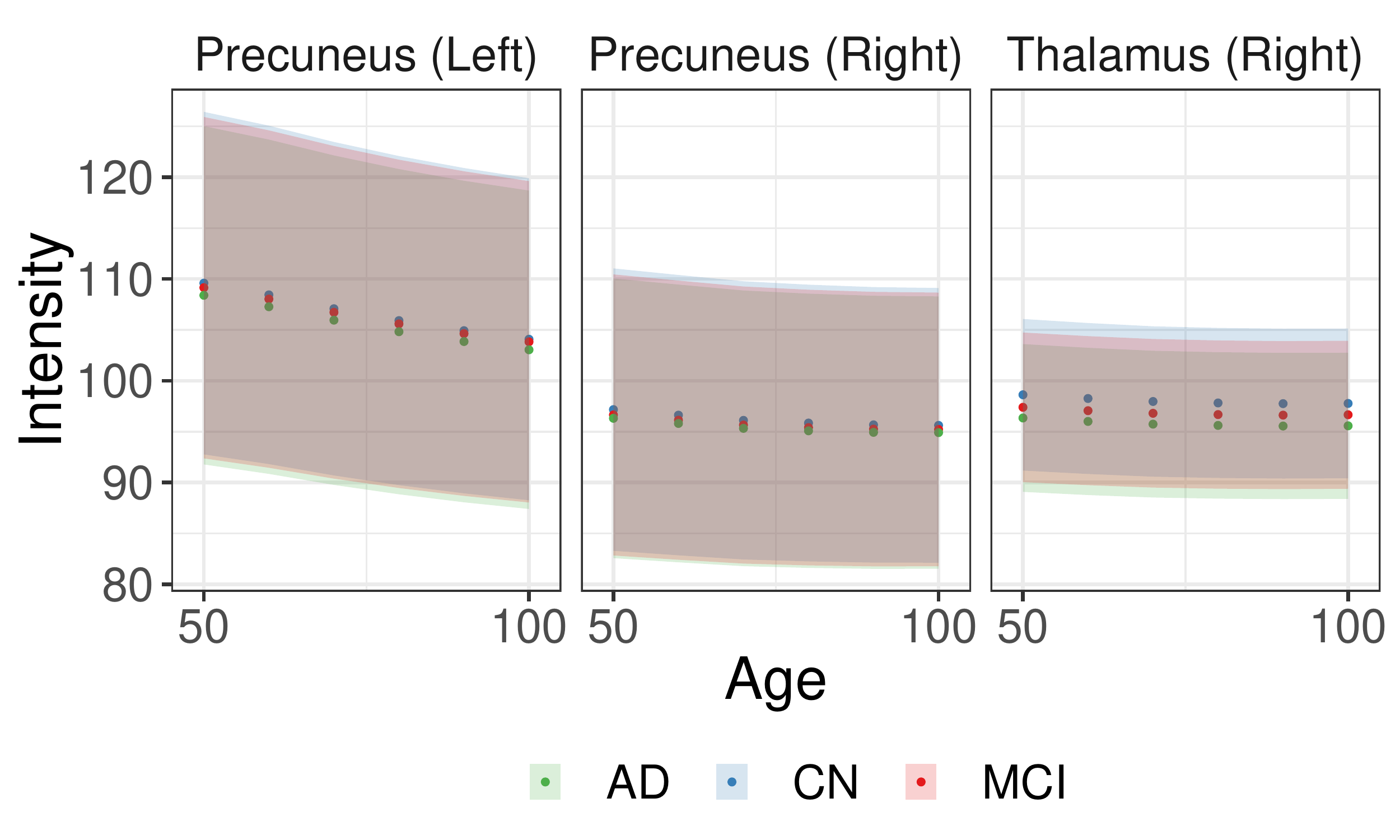}
	\vspace{-7pt}
	\caption{\label{fig:cond_quan_r} The mean intensity with 95\% standard deviation bands of $3$ ROIs with the change of age for all test subjects. The clear downward trend reflects expected hypometabolism as a function of age.}
	\vspace{-10pt}
\end{figure}

\noindent{\bf Adjusting for Age with Conditioning.}
The conditional framework naturally allows us to evaluate potential developing pathology as an individual ages. Training the
full conditional DUAL-GLOW model, we use ground truth ``side'' information (age) as the conditioning variable described above. Figure \ref{fig:cond_samples} shows the continuous change in the $3$ generated images given various age labels for the same MRI. The original image (left) is at age $50$, and as we increase age from $60$ to $100$, increased hypometabolism becomes clear. To quantitatively evaluate our conditional framework, we plot the mean intensity value of a few key ROIs.
As we see in Figure \ref{fig:cond_quan_r}, 
the mean intensity values show a downward trend with age, as expected.
While there is a clear `shift' between AD, MCI, and CN subjects (blue dots lie above red dots, etc.), the wide variance bands indicate a larger sample size may be necessary to
derive statistically sound conclusions (e.g., regarding group differences).
Additional results and details can be found in the appendix.

\subsection{\label{sec: other} Other Potential Applications}
\begin{figure}
	\centering
	\begin{tabular}{c|c|c}
		Input & Generated & Ground Truth \\
		\begin{subfigure}{2.2cm}
			\centering\includegraphics[width=\columnwidth]{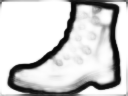}
		\end{subfigure}&
		\begin{subfigure}{2.2cm}
			\centering\includegraphics[width=\columnwidth]{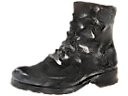}
		\end{subfigure}&
		\begin{subfigure}{2.2cm}
			\centering\includegraphics[width=\columnwidth]{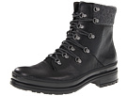}
		\end{subfigure}\\ 
		& & \\
		
		\begin{subfigure}{2.2cm}
			\centering\includegraphics[width=\columnwidth]{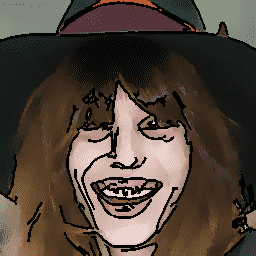}
		\end{subfigure}&
		\begin{subfigure}{2.2cm}
			\centering\includegraphics[width=\columnwidth]{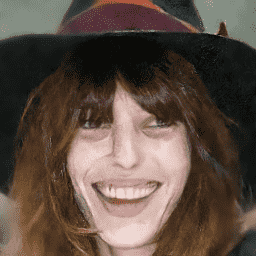}
		\end{subfigure}&
		\begin{subfigure}{2.2cm}
			\centering\includegraphics[width=\columnwidth]{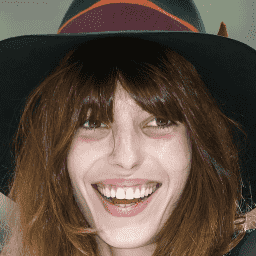}
		\end{subfigure}
	\end{tabular}
	\vspace{-5pt}
	\caption{\label{fig:shoe} Sample generation using DUAL-GLOW. The first row: UT-Zap50K dataset. The second row: CartoonFace dataset.}
	\vspace{-14pt}
\end{figure}

\vspace{-5pt}
While not a key focus of our work, to show the model's generality on visually
familiar images we directly test DUAL-GLOW's ability to generate images on a standard computer
vision modality transfer task. Using the UT-Zap50K dataset \cite{semjitter,finegrained} of shoe images,
we construct HED \cite{xie2015holistically} edge images as ``sketches", similar to \cite{isola2017image}.
We aim to learn a mapping from sketch to shoe.
We also create a cartoon face dataset based on CelebA \cite{liu2015deep} and train our model to generate
a realistic image from the cartoon face. Fig. \ref{fig:shoe} shows the results of applying our model (and ground truth).
Clearly, more specialized networks designed for such a task will yield more striking results, but these experiments
suggest that the framework is general and applicable in additional settings. These
results are available on the project homepage and in the appendix.

\section{Conclusions}
\vspace{-7pt}
We propose a flow-based generative model, DUAL-GLOW,
for inter-modality transformation in medical imaging.
The model allows for end-to-end PET image generation from MRI
for full three-dimensional volumes, and takes advantage of explicitly characterizing the conditional distribution of one modality given the other.
While inter-modality transfer has been reported using
GANs, we present improved results along with the ability to condition
the output easily. 
Applied to the ADNI dataset,
we are able to generate sharp synthetic PET images that are scientifically meaningful. 
Standard correlation and classification analysis demonstrates the potential
of generated PET in diagnosing Alzheimer's Disease,
and the conditional side information framework is promising
for assessing the change of spatial metabolism with age.

\noindent
\\\large\textbf{Acknowledgments.}
\normalsize
HS was supported by
Natural Science Foundation of China (Grant No. 61876098, 61573219)
and scholarships from China Scholarship Council (CSC). Research
was also supported in part by R01 AG040396, R01 EB022883,
NSF CAREER award RI 1252725, UW Draper Technology Innovation Fund (TIF) award,
UW CPCP AI117924 and a NIH predoctoral fellowship to RM via T32 LM012413.

{\small
\bibliographystyle{ieee_fullname}
\bibliography{refs}
}

\newpage

\section{Appendix}

In this supplement, We provide extensive additional details regarding the neuroimaging experiments and some theoretical analysis.

\subsection{The ADNI dataset}
Data used in the experiments for this work were obtained directly from the Alzheimer’s  Disease Neuroimaging  Initiative  (ADNI)  database (\url{adni.loni.usc.edu}). As such, the investigators within the ADNI contributed to the design and implementation of ADNI and/or provided data but did not participate in analysis or writing of this report. A complete listing of ADNI investigators can be found at: \url{http://adni.loni.usc.edu/wp-content/uploads/how_to_apply/ADNI_Acknowledgement_List.pdf}. The ADNI was launched in 2003 as a public-private partnership, led by Principal Investigator Michael W. Weiner, MD. The primary goal of ADNI has been to test whether serial magnetic resonance imaging (MRI), positron emission tomography (PET), other biologicalmarkers, and  clinical and neuropsychological assessment can be combined to measure the progression of mild cognitive impairment (MCI) and early Alzheimer’s disease (AD).For up-to-date information, see \url{www.adni-info.org}.

Pre-processing details are described in the main paper. MR images were segmented such that the skull and other bone matter were masked, leaving only grey matter, white matter, and cerebrospinal fluid (CSF).
After pre-processing, we obtain $806$ clean MRI/PET pairs. The demographics of the final dataset are shown in Table \ref{demographics}.
As voxel size was fixed for all volumes to $1.5mm^3$, processed images were of dimension $121\times 145\times 121$. Images were cropped and downsampled slightly after skull extraction to allow for faster training.

\begin{table*}[]
	\begin{center}
		\caption{\label{demographics} Demographic details of the full ADNI dataset in our experiments.}
		\begin{tabular}{l  c    c c ccc}
			\hline
			\footnotesize{CATEGORY}     & CN    &SMC  &EMCI &MCI&LMCI &AD \\
			\hline                                                  
			\# of subjects          &$ 259$ &   $18$ & $88$ &  $263$ & $64$ & $114$ \\
			Age (mean)                &$76.47$ & $70.94$ & $71.64$ & $77.86$ & $73.14$ & $75.98$ \\
			Age (std)                &$5.26$ & $4.76$ & $6.55$ & $7.42$ & $6.43$ & $7.27$ \\
			Gender (F/M)             &$116/143$ & $11/7$ & $38/50$ & $66/197$ & $24/40$ & $42/72$ \\
			
			\hline
		\end{tabular}
	\end{center}
\end{table*}

\subsection{Architecture Details}

{\bf ADNI Brain Imaging Experiments.}
There are $4$ ``levels" in our two invertible networks, each ``level" containing $16$ affine coupling layers. The small network with three 3D convolutional layers is shared by two nonlinear operators $s(\cdot)$ and $t(\cdot)$. There are $512$ channels in the intermediate layers. For the hierarchical correction learning network, we split the hidden codes of the output of the first three modules in the invertible network and design four 3D CNNs (relation networks) with $1$ convolutional layer for all latent codes. For the conditional framework case, we concatenate the five discriminators with the adaptive number of layers to the tail of all four levels of the MRI inference network and the top-most level of the PET inference network. 

For other compasion methods, we have $13$ resBlocks of $34$ \textit{convs} with the U-net architecture for C-VAE, $12$ \textit{convs} in \textit{G} and $8$ resBlocks in \textit{D} for the cGAN, $15$ \textit{convs} in \textit{G} and $8$ resBlocks in \textit{D} for UcGAN. Flow-based methods for image translation do not currently exist. But we implemented an iterative Glow (iGlow:\,4 levels,\,$4\times 10$\,coupling layers), concatenating paired MRI and PET as input. After training, we fix networks and set input PET as trainable variables and obtain PET by iteratively optimizing the log-likelihood w.r.t. these variables. But the unstable gradient ascent gives bad results.

{\bf Natural Image Experiments.}
For natural image experiments, the settings of the hypersparameters are shown in Table \ref{archs}. The depth is equal to the number of coupling layers in each "level". Since the resolution of images in UT-Zap50K is $128 \times 128$, we use the "level" of $5$. For the celebA dataset (resolution: $256 \times 256$), the number of the "level" is $6$. The relation network is a CNNs with $8$ conv layers in both two experiments.

\begin{table}[]
	\begin{center}
		\caption{\label{archs} Main hyperparameters in our natural image experiments.}
		\begin{tabular}{l  c    c c }
			\hline
			\footnotesize{DATASET}     & Levels    &Depth   & Layers (Relation Nets)\\
			\hline                                                  
			UT-Zap50K           &$ 5 $  &$ 8 $  &$ 8 $  \\
			Cartoon-Celeba      &$ 6 $  &$ 8 $  &$ 8 $  \\
			\hline
		\end{tabular}
	\end{center}
\end{table}

\subsection{Generated Samples}
\begin{figure*}[]
	\centering
	\begin{tabular}{ccc|ccc}
		MRI & GT & Synthetic & MRI & GT & Synthetic \\
		
		\begin{subfigure}{2.cm}
			\centering\includegraphics[width=2.cm]{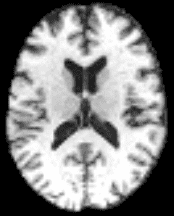}
		\end{subfigure}&
		\begin{subfigure}{2.cm}
			\centering\includegraphics[width=2.cm]{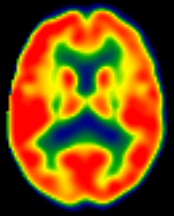}
		\end{subfigure}&
		\begin{subfigure}{2.cm}
			\centering\includegraphics[width=2.cm]{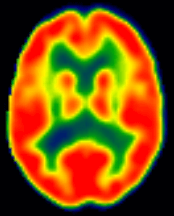}
		\end{subfigure}&
		
		\begin{subfigure}{2.cm}
			\centering\includegraphics[width=2.cm]{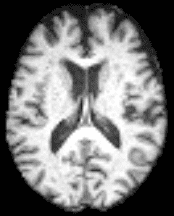}
		\end{subfigure}&
		\begin{subfigure}{2.cm}
			\centering\includegraphics[width=2.cm]{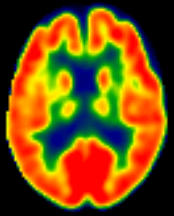}
		\end{subfigure}&
		\begin{subfigure}{2.cm}
			\centering\includegraphics[width=2.cm]{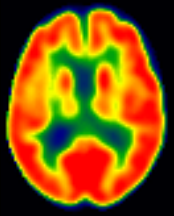}
		\end{subfigure} \\

		\begin{subfigure}{2.cm}
			\centering\includegraphics[width=2.cm]{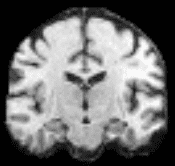}
		\end{subfigure}&
		\begin{subfigure}{2.cm}
			\centering\includegraphics[width=2.cm]{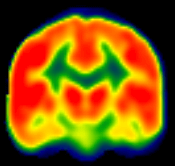}
		\end{subfigure}&
		\begin{subfigure}{2.cm}
			\centering\includegraphics[width=2.cm]{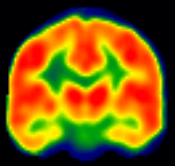}
		\end{subfigure}&
		
		\begin{subfigure}{2.cm}
			\centering\includegraphics[width=2.cm]{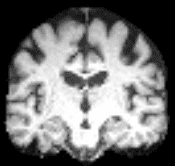}
		\end{subfigure}&
		\begin{subfigure}{2.cm}
			\centering\includegraphics[width=2.cm]{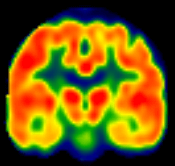}
		\end{subfigure}&
		\begin{subfigure}{2.cm}
			\centering\includegraphics[width=2.cm]{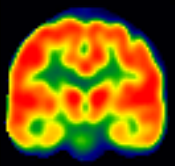}
		\end{subfigure} \\

		\begin{subfigure}{2.cm}
			\centering\includegraphics[width=2.cm]{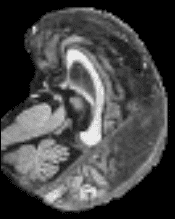}
		\end{subfigure}&
		\begin{subfigure}{2.cm}
			\centering\includegraphics[width=2.cm]{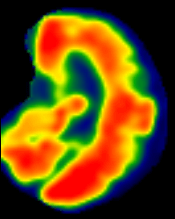}
		\end{subfigure}&
		\begin{subfigure}{2.cm}
			\centering\includegraphics[width=2.cm]{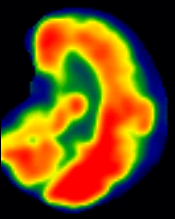}
		\end{subfigure}&
		
		\begin{subfigure}{2.cm}
			\centering\includegraphics[width=2.cm]{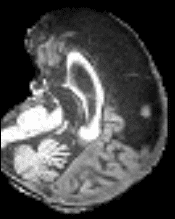}
		\end{subfigure}&
		\begin{subfigure}{2.cm}
			\centering\includegraphics[width=2.cm]{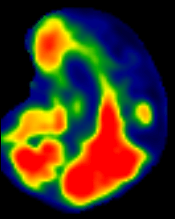}
		\end{subfigure}&
		\begin{subfigure}{2.cm}
			\centering\includegraphics[width=2.cm]{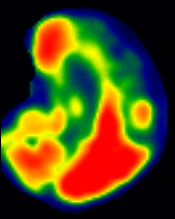}
		\end{subfigure} \\
		
	\end{tabular}

	\caption{\label{fig:samples-mci} Two synthetic MCI subjects. From left to right: input of MRI, the ground truth, the synthetic subject.}
\end{figure*}

\begin{figure*}[]
	\centering
	\begin{tabular}{ccc|ccc}
		MRI & GT & Synthetic & MRI & GT & Synthetic \\
		
		\begin{subfigure}{2.cm}
			\centering\includegraphics[width=2.cm]{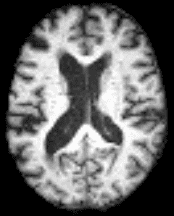}
		\end{subfigure}&
		\begin{subfigure}{2.cm}
			\centering\includegraphics[width=2.cm]{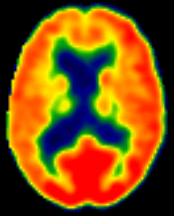}
		\end{subfigure}&
		\begin{subfigure}{2.cm}
			\centering\includegraphics[width=2.cm]{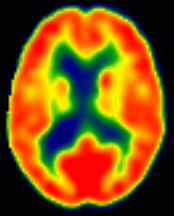}
		\end{subfigure}&
		
		\begin{subfigure}{2.cm}
			\centering\includegraphics[width=2.cm]{4_1_s.png}
		\end{subfigure}&
		\begin{subfigure}{2.cm}
			\centering\includegraphics[width=2.cm]{4_2_s.png}
		\end{subfigure}&
		\begin{subfigure}{2.cm}
			\centering\includegraphics[width=2.cm]{4_3_s.png}
		\end{subfigure} \\

		\begin{subfigure}{2.cm}
			\centering\includegraphics[width=2.cm]{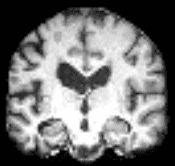}
		\end{subfigure}&
		\begin{subfigure}{2.cm}
			\centering\includegraphics[width=2.cm]{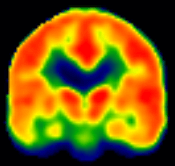}
		\end{subfigure}&
		\begin{subfigure}{2.cm}
			\centering\includegraphics[width=2.cm]{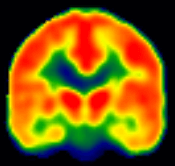}
		\end{subfigure}&
		
		\begin{subfigure}{2.cm}
			\centering\includegraphics[width=2.cm]{4_1_a.png}
		\end{subfigure}&
		\begin{subfigure}{2.cm}
			\centering\includegraphics[width=2.cm]{4_2_a.png}
		\end{subfigure}&
		\begin{subfigure}{2.cm}
			\centering\includegraphics[width=2.cm]{4_3_a.png}
		\end{subfigure} \\

		\begin{subfigure}{2.cm}
			\centering\includegraphics[width=2.cm]{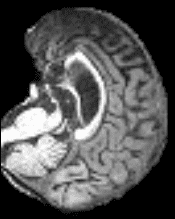}
		\end{subfigure}&
		\begin{subfigure}{2.cm}
			\centering\includegraphics[width=2.cm]{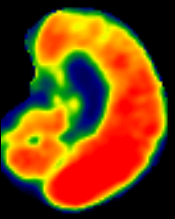}
		\end{subfigure}&
		\begin{subfigure}{2.cm}
			\centering\includegraphics[width=2.cm]{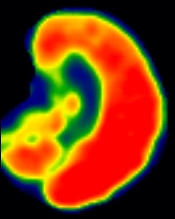}
		\end{subfigure}&
		
		\begin{subfigure}{2.cm}
			\centering\includegraphics[width=2.cm]{4_1_c.png}
		\end{subfigure}&
		\begin{subfigure}{2.cm}
			\centering\includegraphics[width=2.cm]{4_2_c.png}
		\end{subfigure}&
		\begin{subfigure}{2.cm}
			\centering\includegraphics[width=2.cm]{4_3_c.png}
		\end{subfigure} \\
		
	\end{tabular}

	\caption{\label{fig:samples-ad} Two synthetic AD subjects. From left to right: input of MRI, the ground truth, the synthetic subject.}
\end{figure*}

{\bf Brain Imaging Experiments.}
The images that follow are additional representative samples from our framework. Figures \ref{fig:samples-cn}, \ref{fig:samples-mci}, \ref{fig:samples-ad} show ground truth and reconstructions of cognitively healthy, mildly cognitively impaired, and Alzheimer's diseased patients within our test group.

\begin{figure*}[]
	\centering
	\includegraphics[width=1.95\columnwidth]{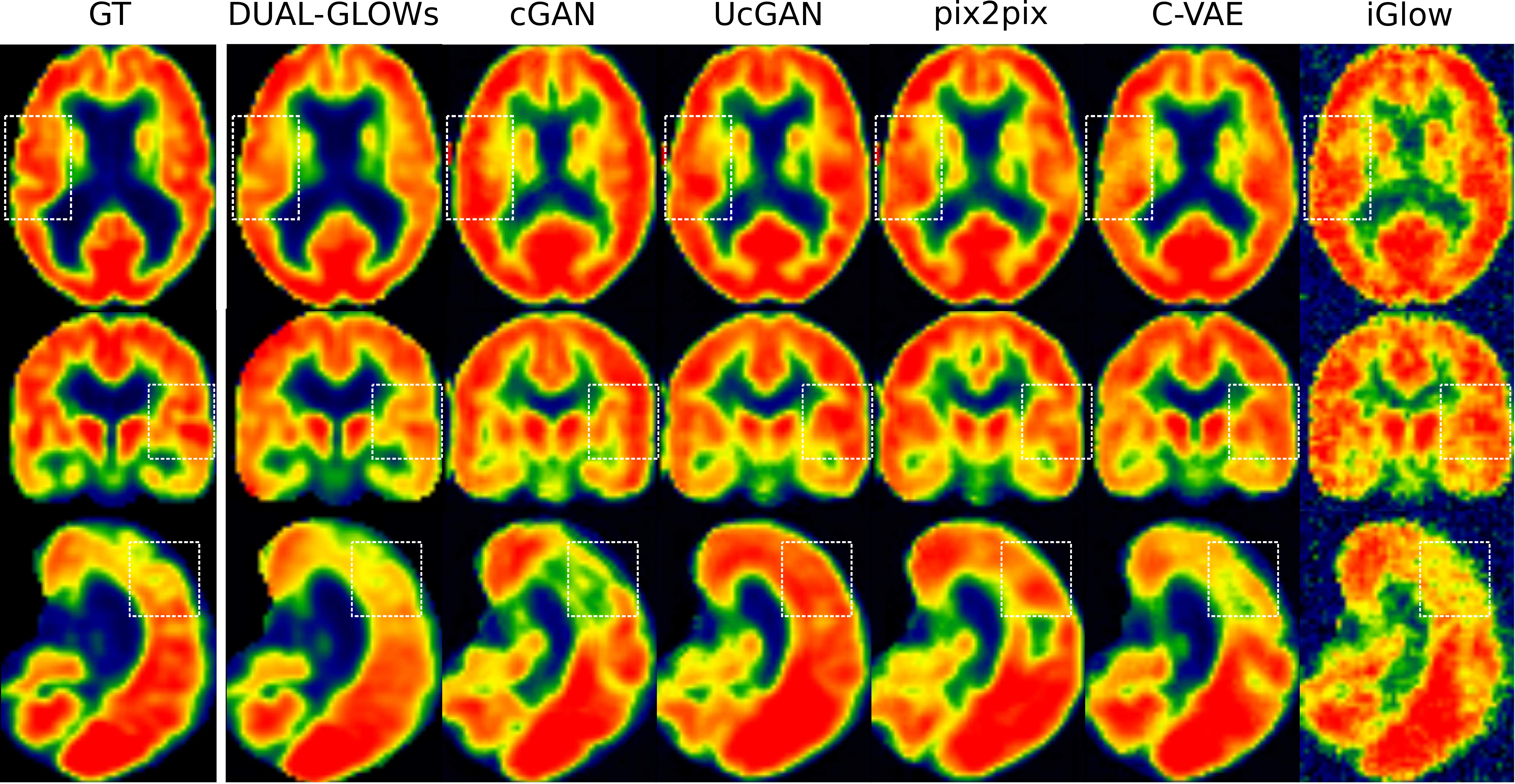}
	\caption{\label{fig:comparison} DUAL-GLOW can produce more accurate prediction in dashed rectangles.}
\end{figure*}

Figure \ref{fig:comparison} shows the comparison visualization results, which shows that DUAL-GLOW outperforms other methods in most regions.

\begin{figure*}[]
	\centering
	\includegraphics[width=1.95\columnwidth]{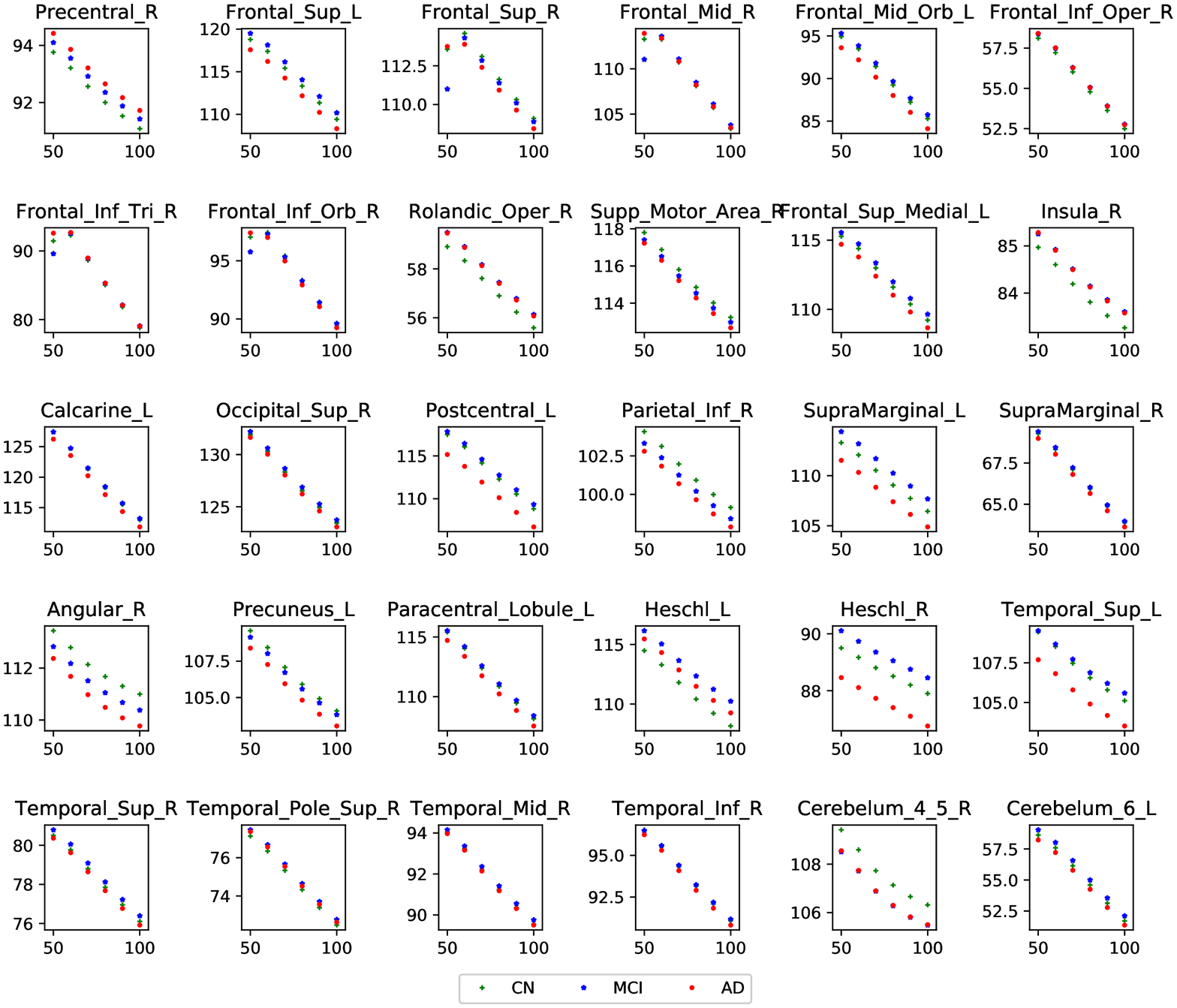}
	\caption{\label{fig:cond} Decreasing trends for $30$ ROIs (related to aging).}
\end{figure*}

Figure \ref{fig:cond_samples} shows additional age conditioning results, again for each of the three disease groups (CN, MCI, and AD). We also plot the mean intensity of other $30$ ROIs of $3$ subjects given $6$ age labels (from 50 to 100) in Fig \ref{fig:cond}, which shows a clear decreasing trend, i.e., decreased metabolism with aging.


\begin{figure*}[b]
	\centering
	\begin{tabular}{ccc|ccc}
		MRI & GT & Synthetic & MRI & GT & Synthetic \\
		
		\begin{subfigure}{2.cm}
			\centering\includegraphics[width=2.cm]{2_1_s.png}
		\end{subfigure}&
		\begin{subfigure}{2.cm}
			\centering\includegraphics[width=2.cm]{2_2_s.png}
		\end{subfigure}&
		\begin{subfigure}{2.cm}
			\centering\includegraphics[width=2.cm]{2_3_s.png}
		\end{subfigure}&
		
		\begin{subfigure}{2.cm}
			\centering\includegraphics[width=2.cm]{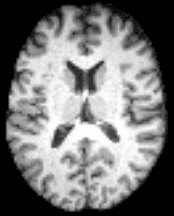}
		\end{subfigure}&
		\begin{subfigure}{2.cm}
			\centering\includegraphics[width=2.cm]{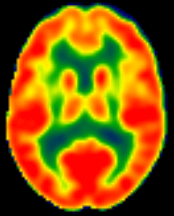}
		\end{subfigure}&
		\begin{subfigure}{2.cm}
			\centering\includegraphics[width=2.cm]{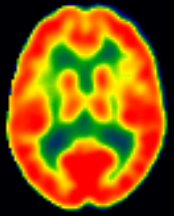}
		\end{subfigure} \\

		\begin{subfigure}{2.cm}
			\centering\includegraphics[width=2.cm]{2_1_a.png}
		\end{subfigure}&
		\begin{subfigure}{2.cm}
			\centering\includegraphics[width=2.cm]{2_2_a.png}
		\end{subfigure}&
		\begin{subfigure}{2.cm}
			\centering\includegraphics[width=2.cm]{2_3_a.png}
		\end{subfigure}&
		
		\begin{subfigure}{2.cm}
			\centering\includegraphics[width=2.cm]{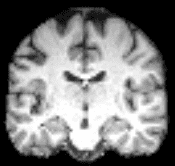}
		\end{subfigure}&
		\begin{subfigure}{2.cm}
			\centering\includegraphics[width=2.cm]{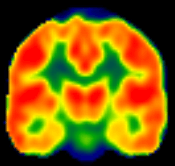}
		\end{subfigure}&
		\begin{subfigure}{2.cm}
			\centering\includegraphics[width=2.cm]{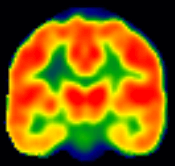}
		\end{subfigure} \\

		\begin{subfigure}{2.cm}
			\centering\includegraphics[width=2.cm]{2_1_c.png}
		\end{subfigure}&
		\begin{subfigure}{2.cm}
			\centering\includegraphics[width=2.cm]{2_2_c.png}
		\end{subfigure}&
		\begin{subfigure}{2.cm}
			\centering\includegraphics[width=2.cm]{2_3_c.png}
		\end{subfigure}&
		
		\begin{subfigure}{2.cm}
			\centering\includegraphics[width=2.cm]{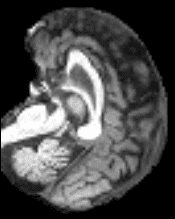}
		\end{subfigure}&
		\begin{subfigure}{2.cm}
			\centering\includegraphics[width=2.cm]{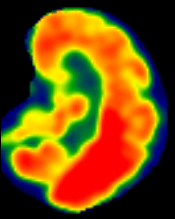}
		\end{subfigure}&
		\begin{subfigure}{2.cm}
			\centering\includegraphics[width=2.cm]{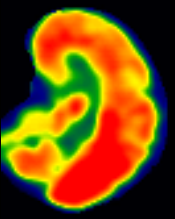}
		\end{subfigure} \\
		
	\end{tabular}

	\caption{\label{fig:samples-cn} Two synthetic CN subjects. From left to right: input of MRI, the ground truth, the synthetic subject.}
\end{figure*}

\begin{figure*}[]
	\centering
	\vspace{-8mm}
	\begin{tabular}{C{3mm} ccccccc}
		& MRI & 50 & 60 & 70 & 80 & 90 & 100 \\
		&
		\begin{subfigure}{1.5cm}
			\centering\includegraphics[width=1.5cm]{3_7_s_cond.png}
		\end{subfigure} &
		\begin{subfigure}{1.5cm}
			\centering\includegraphics[width=1.5cm]{3_1_s_cond.png}
		\end{subfigure}&
		\begin{subfigure}{1.5cm}
			\centering\includegraphics[width=1.5cm]{3_2_s_cond.png}
		\end{subfigure}&
		\begin{subfigure}{1.5cm}
			\centering\includegraphics[width=1.5cm]{3_3_s_cond.png}
		\end{subfigure}&
		
		\begin{subfigure}{1.5cm}
			\centering\includegraphics[width=1.5cm]{3_4_s_cond.png}
		\end{subfigure}&
		\begin{subfigure}{1.5cm}
			\centering\includegraphics[width=1.5cm]{3_5_s_cond.png}
		\end{subfigure}&
		\begin{subfigure}{1.5cm}
			\centering\includegraphics[width=1.5cm]{3_6_s_cond.png}
		\end{subfigure}\\

		\multirow{3}{*}{AD}&
		\begin{subfigure}{1.5cm}
			\centering\includegraphics[width=1.5cm]{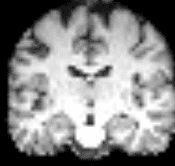}
		\end{subfigure} &
		\begin{subfigure}{1.5cm}
			\centering\includegraphics[width=1.5cm]{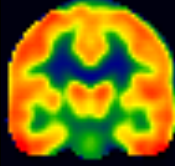}
		\end{subfigure}&
		\begin{subfigure}{1.5cm}
			\centering\includegraphics[width=1.5cm]{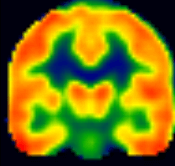}
		\end{subfigure}&
		\begin{subfigure}{1.5cm}
			\centering\includegraphics[width=1.5cm]{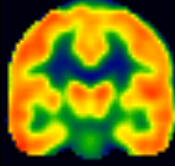}
		\end{subfigure}&
		
		\begin{subfigure}{1.5cm}
			\centering\includegraphics[width=1.5cm]{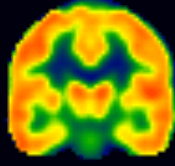}
		\end{subfigure}&
		\begin{subfigure}{1.5cm}
			\centering\includegraphics[width=1.5cm]{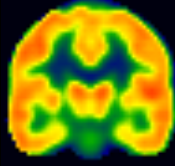}
		\end{subfigure}&
		\begin{subfigure}{1.5cm}
			\centering\includegraphics[width=1.5cm]{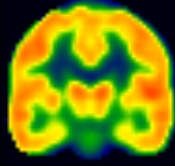}
		\end{subfigure}\\

		&
		\begin{subfigure}{1.5cm}
			\centering\includegraphics[width=1.5cm]{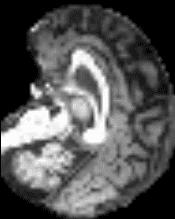}
		\end{subfigure} &
		\begin{subfigure}{1.5cm}
			\centering\includegraphics[width=1.5cm]{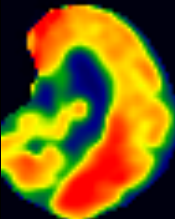}
		\end{subfigure}&
		\begin{subfigure}{1.5cm}
			\centering\includegraphics[width=1.5cm]{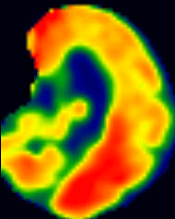}
		\end{subfigure}&
		\begin{subfigure}{1.5cm}
			\centering\includegraphics[width=1.5cm]{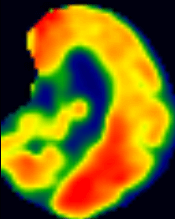}
		\end{subfigure}&
		
		\begin{subfigure}{1.5cm}
			\centering\includegraphics[width=1.5cm]{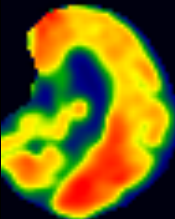}
		\end{subfigure}&
		\begin{subfigure}{1.5cm}
			\centering\includegraphics[width=1.5cm]{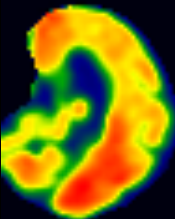}
		\end{subfigure}&
		\begin{subfigure}{1.5cm}
			\centering\includegraphics[width=1.5cm]{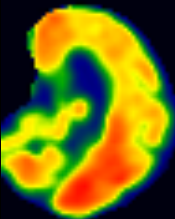}
		\end{subfigure}\\
		
		\vspace{-3mm}
		\\
		
		\hline
		\vspace{-3mm}
		\\
		
		&
		\begin{subfigure}{1.5cm}
			\centering\includegraphics[width=1.5cm]{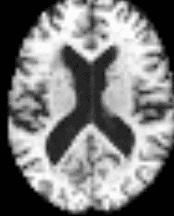}
		\end{subfigure}&
		\begin{subfigure}{1.5cm}
			\centering\includegraphics[width=1.5cm]{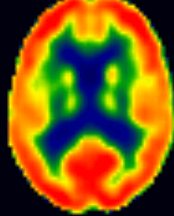}
		\end{subfigure}&
		\begin{subfigure}{1.5cm}
			\centering\includegraphics[width=1.5cm]{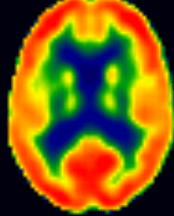}
		\end{subfigure}&
		\begin{subfigure}{1.5cm}
			\centering\includegraphics[width=1.5cm]{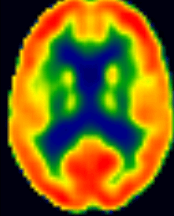}
		\end{subfigure}&
		
		\begin{subfigure}{1.5cm}
			\centering\includegraphics[width=1.5cm]{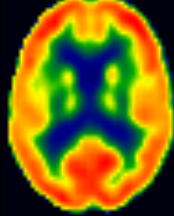}
		\end{subfigure}&
		\begin{subfigure}{1.5cm}
			\centering\includegraphics[width=1.5cm]{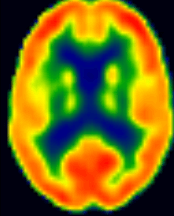}
		\end{subfigure}&
		\begin{subfigure}{1.5cm}
			\centering\includegraphics[width=1.5cm]{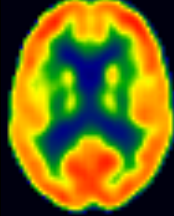}
		\end{subfigure} \\

		\multirow{3}{*}{MCI}&
		\begin{subfigure}{1.5cm}
			\centering\includegraphics[width=1.5cm]{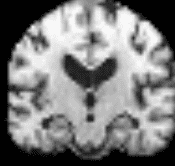}
		\end{subfigure} &
		\begin{subfigure}{1.5cm}
			\centering\includegraphics[width=1.5cm]{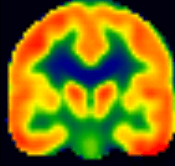}
		\end{subfigure}&
		\begin{subfigure}{1.5cm}
			\centering\includegraphics[width=1.5cm]{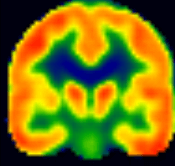}
		\end{subfigure}&
		\begin{subfigure}{1.5cm}
			\centering\includegraphics[width=1.5cm]{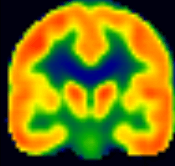}
		\end{subfigure}&
		
		\begin{subfigure}{1.5cm}
			\centering\includegraphics[width=1.5cm]{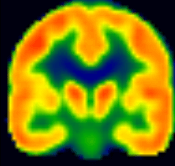}
		\end{subfigure}&
		\begin{subfigure}{1.5cm}
			\centering\includegraphics[width=1.5cm]{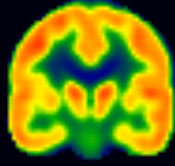}
		\end{subfigure}&
		\begin{subfigure}{1.5cm}
			\centering\includegraphics[width=1.5cm]{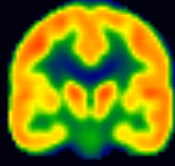}
		\end{subfigure}\\

		&
		\begin{subfigure}{1.5cm}
			\centering\includegraphics[width=1.5cm]{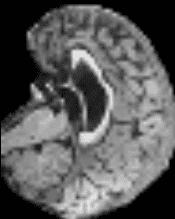}
		\end{subfigure} &
		\begin{subfigure}{1.5cm}
			\centering\includegraphics[width=1.5cm]{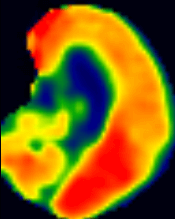}
		\end{subfigure}&
		\begin{subfigure}{1.5cm}
			\centering\includegraphics[width=1.5cm]{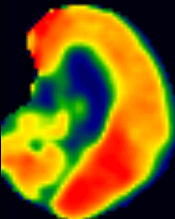}
		\end{subfigure}&
		\begin{subfigure}{1.5cm}
			\centering\includegraphics[width=1.5cm]{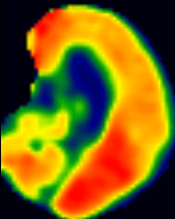}
		\end{subfigure}&
		
		\begin{subfigure}{1.5cm}
			\centering\includegraphics[width=1.5cm]{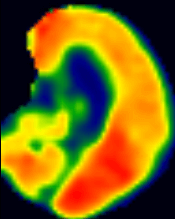}
		\end{subfigure}&
		\begin{subfigure}{1.5cm}
			\centering\includegraphics[width=1.5cm]{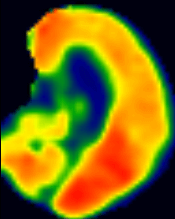}
		\end{subfigure}&
		\begin{subfigure}{1.5cm}
			\centering\includegraphics[width=1.5cm]{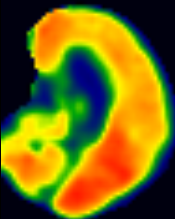}
		\end{subfigure}\\

		\vspace{-3mm}
		\\
		
		\hline
		\vspace{-3mm}
		\\
		&
		\begin{subfigure}{1.5cm}
			\centering\includegraphics[width=1.5cm]{2_7_s_cond.png}
		\end{subfigure} &
		\begin{subfigure}{1.5cm}
			\centering\includegraphics[width=1.5cm]{2_1_s_cond.png}
		\end{subfigure}&
		\begin{subfigure}{1.5cm}
			\centering\includegraphics[width=1.5cm]{2_2_s_cond.png}
		\end{subfigure}&
		\begin{subfigure}{1.5cm}
			\centering\includegraphics[width=1.5cm]{2_3_s_cond.png}
		\end{subfigure}&
		
		\begin{subfigure}{1.5cm}
			\centering\includegraphics[width=1.5cm]{2_4_s_cond.png}
		\end{subfigure}&
		\begin{subfigure}{1.5cm}
			\centering\includegraphics[width=1.5cm]{2_5_s_cond.png}
		\end{subfigure}&
		\begin{subfigure}{1.5cm}
			\centering\includegraphics[width=1.5cm]{2_6_s_cond.png}
		\end{subfigure}\\

		\multirow{3}{*}{CN}&
		\begin{subfigure}{1.5cm}
			\centering\includegraphics[width=1.5cm]{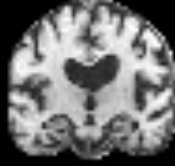}
		\end{subfigure} &
		\begin{subfigure}{1.5cm}
			\centering\includegraphics[width=1.5cm]{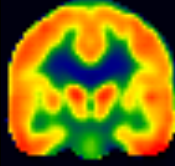}
		\end{subfigure}&
		\begin{subfigure}{1.5cm}
			\centering\includegraphics[width=1.5cm]{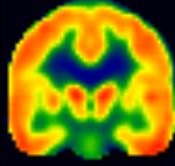}
		\end{subfigure}&
		\begin{subfigure}{1.5cm}
			\centering\includegraphics[width=1.5cm]{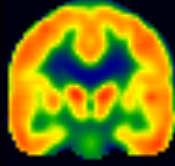}
		\end{subfigure}&
		
		\begin{subfigure}{1.5cm}
			\centering\includegraphics[width=1.5cm]{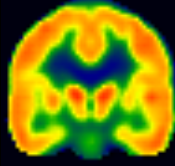}
		\end{subfigure}&
		\begin{subfigure}{1.5cm}
			\centering\includegraphics[width=1.5cm]{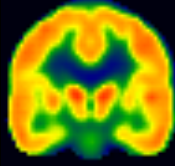}
		\end{subfigure}&
		\begin{subfigure}{1.5cm}
			\centering\includegraphics[width=1.5cm]{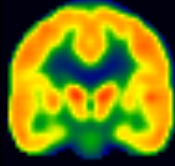}
		\end{subfigure}\\

		&
		\begin{subfigure}{1.5cm}
			\centering\includegraphics[width=1.5cm]{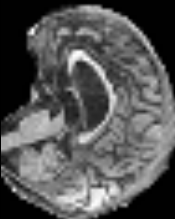}
		\end{subfigure} &
		\begin{subfigure}{1.5cm}
			\centering\includegraphics[width=1.5cm]{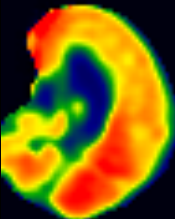}
		\end{subfigure}&
		\begin{subfigure}{1.5cm}
			\centering\includegraphics[width=1.5cm]{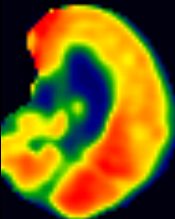}
		\end{subfigure}&
		\begin{subfigure}{1.5cm}
			\centering\includegraphics[width=1.5cm]{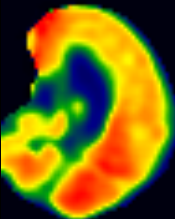}
		\end{subfigure}&
		
		\begin{subfigure}{1.5cm}
			\centering\includegraphics[width=1.5cm]{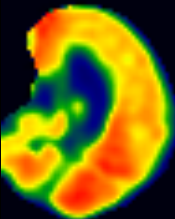}
		\end{subfigure}&
		\begin{subfigure}{1.5cm}
			\centering\includegraphics[width=1.5cm]{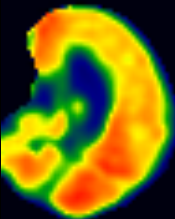}
		\end{subfigure}&
		\begin{subfigure}{1.5cm}
			\centering\includegraphics[width=1.5cm]{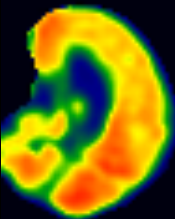}
		\end{subfigure}\\
		
	\end{tabular}

	\caption{\label{fig:cond_samples} Age information manipulation. There are 3 subjects, AD, MCI, and CN, each subject provides $6$ results \textit{w.r.t.} a variant of age labels. As we scan left to right, we indeed see a decrease trend in metabolism (less red, more yellow) which is completely consistent with what we would expect in aging.}
\end{figure*}

\begin{figure*}[b]
	\centering
	\includegraphics[width=2.1\columnwidth]{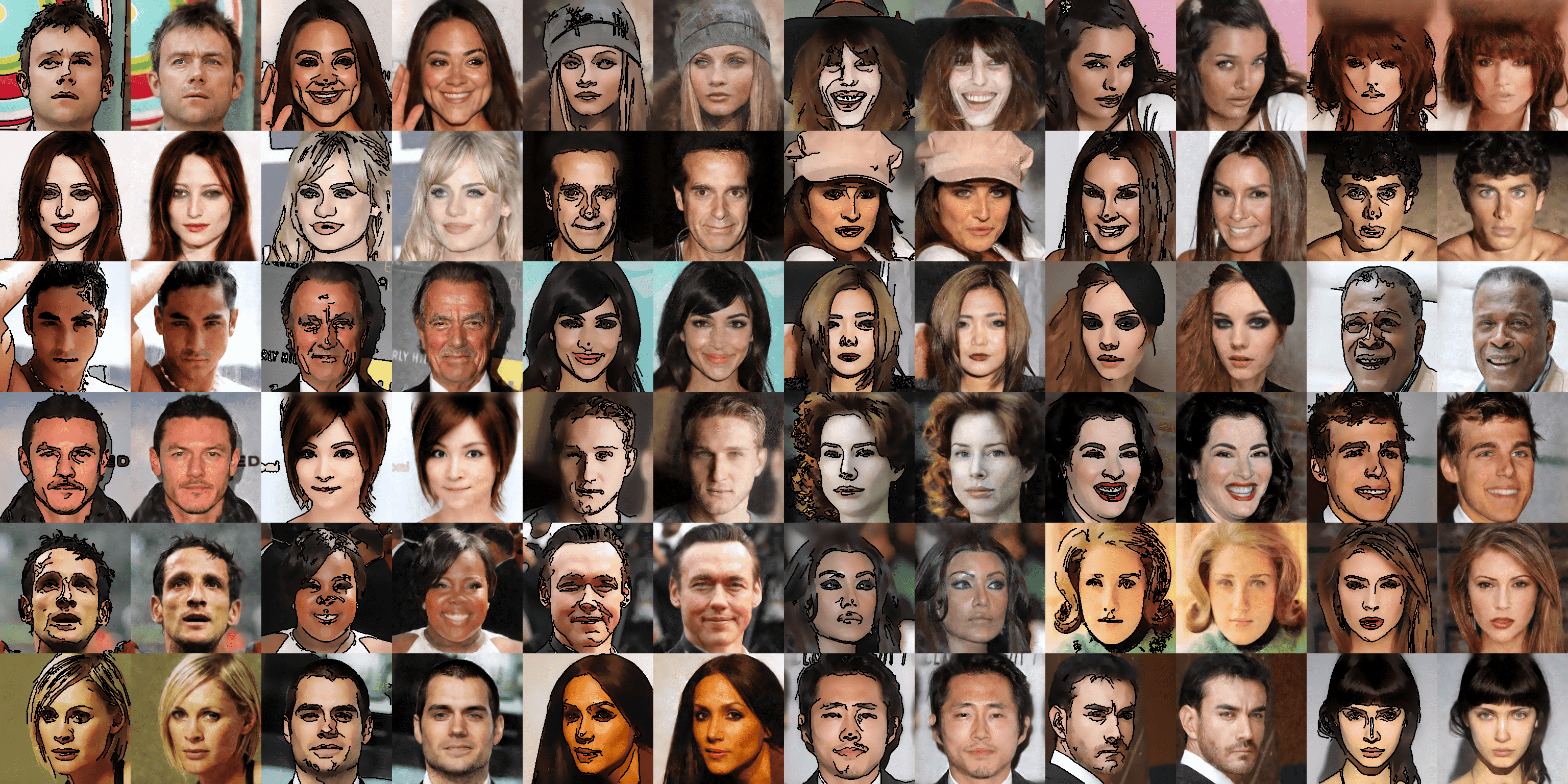}
	\caption{\label{fig:celeba}Input cartoon images and generated/reconstructed faces applying our DUAL-GLOW framework to the CelebA dataset.}
\end{figure*}

\begin{figure*}
	\centering
	\includegraphics[width=2.1\columnwidth]{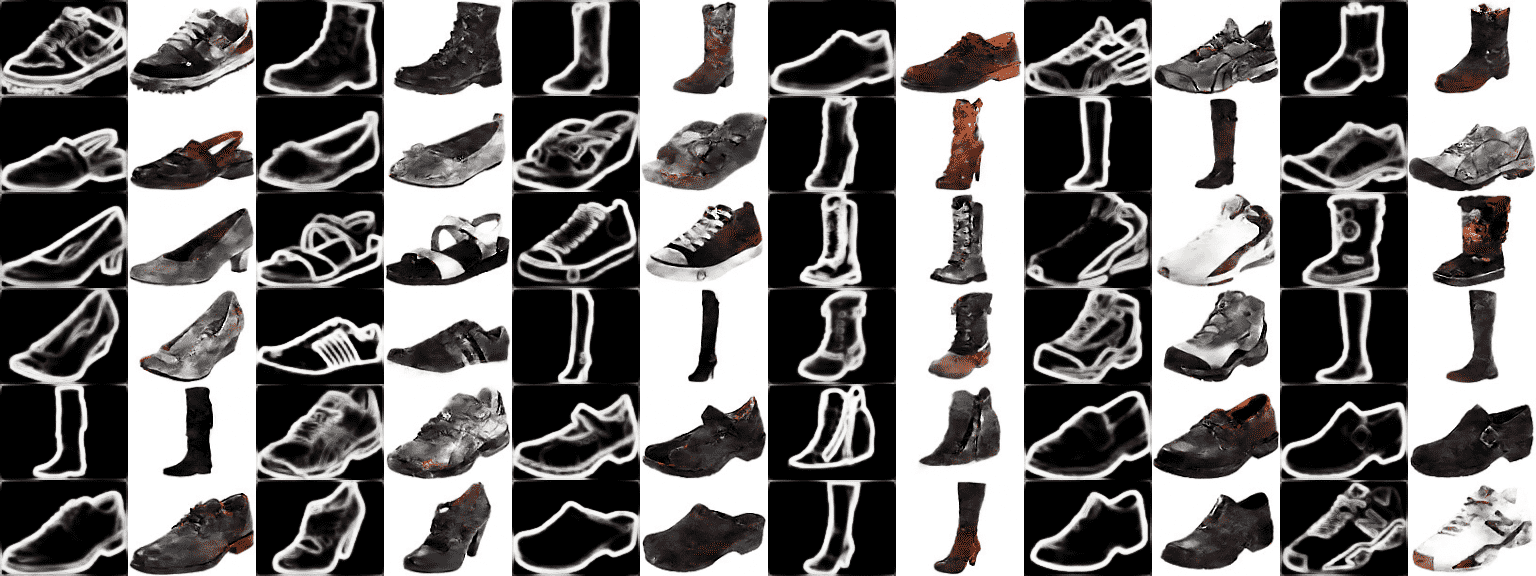}
	\caption{\label{fig:shoes} Input ``sketch" images and generated/reconstructed shoes applying our DUAL-GLOW framework to the UT-Zap50K dataset.}
\end{figure*}

{\bf Other Experiments.}
While not the focus of our work, we show some additional results on two standard imaging datasets: the cartoon-to-face translation in Figure \ref{fig:celeba}; the sketch-to-shoe snynthesis in Figure \ref{fig:shoes}. The input of the CelebA face dataset is the cartoon image processed by using the technique in \textit{opencv}. For UT-Zap50K shoe dataset, we extract edge images as “sketches” by using HED (Holistically-nested edge detection) and learn a mapping from sketch to shoe.

\subsection{Theoretical Analysis}

{\bf Hierarchical architecture.} {
	In Fig\,3/paper, 
	half of the feature map is used as input to the next level. 
	The computational complexity is mainly dependent on the input of each coupling layer. 
	Let $\ell$ denote the number of levels. Supposing the input size is $2^{\ell}$ and the time complexity for each level is $\mathcal{O}(N)$, 
	we have the time complexity of $\mathcal{O}( 2^{\ell} N\ell)$ for the flat architecture and  $\mathcal{O}( (2^{\ell +1} -2) N)$ for the hierarchical one. A larger $\ell$ leads to further reduction.}

\begin{figure}[H]
	\centering
	\includegraphics[width=1.\columnwidth]{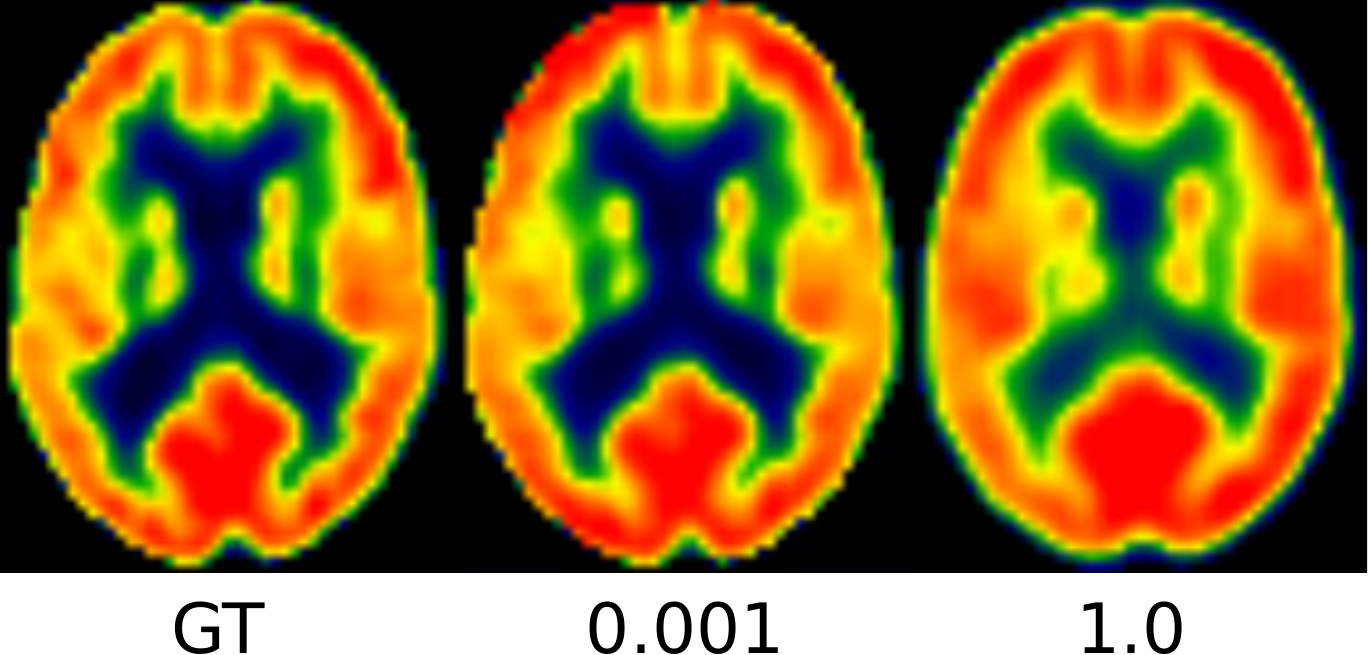}
	\caption{\label{fig:lbd} Better for the small $\lambda$.}
\end{figure}

{\bf The choice of $\lambda$.}
$\lambda$ regularizes networks for MRI. Our goal is to model the conditional distribution rather than the joint distribution. A lower weight on this constraint ($0.001$ in all experiments) leads to easier optimization and better qualitative results (Fig \ref{fig:lbd}).

\end{document}